\documentclass[%
 reprint,
 superscriptaddress,
 amsmath,amssymb,
 aps,
]{revtex4-1}

\usepackage{hyperref}
\hypersetup{
	colorlinks=true, 
	linkcolor=blue,  
	citecolor=cyan,  
}

\usepackage{epstopdf}
\usepackage{graphicx}
\usepackage{dcolumn}
\usepackage{bm}

\begin{document}


\title{Scalar perturbations and quasi-normal modes of a non-linear magnetic-charged black hole surrounded by quintessence}

\author{Hrishikesh~Chakrabarty}
\affiliation{Center for Field Theory and Particle Physics and Department of Physics, Fudan University, 200438 Shanghai, China}
\author{Ahmadjon~Abdujabbarov}
\affiliation{Center for Field Theory and Particle Physics and Department of Physics, Fudan University, 200438 Shanghai, China}
\affiliation{Ulugh Beg Astronomical Institute, Astronomicheskaya 33,
	Tashkent 100052, Uzbekistan }

\author{Cosimo~Bambi}
\email[Corresponding author: ]{bambi@fudan.edu.cn}
\affiliation{Center for Field Theory and Particle Physics and Department of Physics, Fudan University, 200438 Shanghai, China}

\date{\today}

\begin{abstract}

We study scalar perturbations and quasinormal modes of a nonlinear magnetic charged black hole surrounded by quintessence. Time evolution of scalar perturbations is studied for different parameters associated with the black hole solution. We also study the reflection and transmission coefficients along with absorption cross-section for the considered black hole spacetime.  
It was shown that the real part of quasinormal frequency increases with increase in nonlinear magnetic charge while  the module of the imaginary part of the frequency decreases.
The analysis of the perturbations with changing quintessential parameter $c$ showed that perturbations with high values of $ c $ become unstable.

\end{abstract}

\pacs{Valid PACS appear here}
\maketitle


\section{Introduction}\label{intro}

The stability of black hole spacetimes is one of the most interesting questions in general relativity and provides us with many answers related to the black hole itself. The study of various types of perturbations such as scalar, electromagnetic and gravitational on a black hole background is an active area of research. 
Dynamical evolution of any kind of perturbations on a black hole background can be classified into three stages: the initial outburst of the waves, damped oscillation which is also known as quasinormal modes (QNMs), and the late time power law tail. The first stage completely depends upon the initial perturbing field and it does not give us much information about the stability. The second stage is extremely important for black hole stability analysis and it consists of complex frequencies, the real part of which represents the real frequency of the perturbation and the imaginary part represents the damping. These quasinormal modes also provide information about different black hole parameters such as mass, angular momentum, charge etc.

The pioneer work on metric perturbations of the Schwarzschild black hole has been studied by Regge and Wheeler ~\cite{Regge57} and Zerilli~\cite{Zerilli70}.
The author of Ref.~\cite{Vishveshwara70} analyzed numerically scattering of waves on the Schwarzschild black hole~\cite{Nagar05}. Later Chandrasekhar presented a monograph about perturbation theory of black holes~\cite{Chandrasekhar98}. 
The main equation of the perturbation theory is a Shr\"{o}dinger-like equation and, usually, it can be solved using semiclassical or numerical methods. 
Many authors studied this type of perturbative investigations of black holes (see, e.g., Refs.~\cite{Kokkotas99,Berti09a,Konoplya11} and reference therein).

The discovery of gravitational waves opened a new window for black hole perturbation physics~\cite{LIGO16,LIGO16b,Thorne97}. 
It has been discussed that the precision of the observation/experiment of gravitational waves allows us to test alternative theories of gravity~\cite{Konoplya16b,Abramowicz16}. Moreover, it was also shown that, regardless of the existence of horizons, waveforms can be formed~\cite{Cardoso16,Cardoso16erratum}. On the other hand, it was stated that gravastars cannot be formed in a binary black hole merger process~\cite{Chirenti16}. 
There are other works related to the study of the ringdown process through perturbation of black holes ~\cite{Starinets02,Berti03,Kurita02,Vazquez02,Kurita03,Casali04,Maeda05,Seahra05,Abdalla06,Berti06,Kanti06,Park06,Chen07,Starinets09,Nozawa08,Zhidenko08,Cho08,Al-Binni07,Morgan09,Abdalla10,Delsate11,Hod11,Chung16,Janiszewski16,Toshmatov15b,Toshmatov16,Toshmatov17a,Toshmatov17b,Toshmatov18,Toshmatov17c,Stuchlik17}.

The solutions of Einstein equations for black holes have a singularity problem. The appearance of singularities can be considered as a defect of Einstein's general relativity. However, some regular black hole solutions have been obtained by various authors introducing nonlinear electrodynamics to the background gravity~\cite{Ayon-Beato98,Ayon-Beato99,Ayon-Beato99a,Beato00,Bardeen68,Hayward06,Bambi13}.
Different properties of regular black holes have been studied in the literature~\cite{Toshmatov14,Toshmatov15a,Abdujabbarov16b,Abdujabbarov17,Bambi17,Chakrabarty18a}. 

Another interesting subject is the vacuum energy or quintessence, the existence of which changes the structure of the spacetime at asymptotic infinity -- it will be not flat anymore. Furthermore, there will be cosmological horizon and behind that the geometry becomes dynamic. Particularly, 
the effects of a repulsive cosmological constant are widely discussed in~\cite{Stuchlik83,Stuchlik84,Uzan11,Grenon10,Grenon11,Fleury13,Faraoni14,Faraoni15,Stuchlik11,Schee13,Stuchlik12c,Stuchlik99a,Stuchlik02,Stuchlik04,Kraniotis04,Kraniotis05,Kraniotis07,Kagramanova06,Stuchlik00,Slany05,Rezzolla03a,Stuchlik08,Stuchlik09a}. Study of quasinormal modes of these black holes surrounded by quintessence is an interesting topic. The effects of the quintessential parameter on quasinormal frequencies of different black hole spacetimes were studied in~\cite{Zhang07,Zhang09,Graca18,
Chen05,Chen08,Zhang06,Tharanath14,Saleh18,Varghese09}. Recently it has been obtained the solution of a regular black hole in the presence of quintessence~\cite{Nam18}. In this work, we study the stability of this solution and the quasinormal modes of perturbations. We also study the reflection coefficient, greybody factor, and absorption coefficient of scalar perturbations. 

The paper is organized as follows. In Sect.~\ref{bhspacetime}, we review the regular black hole solution surrounded by quintessence. In Sect.~\ref{scalarpert}, we study the basic equations of  scalar perturbations and, in Sect.~\ref{numresult}, we give the numerical results of evolution of scalar perturbations and quasinormal modes. Sect.~\ref{greybody} is devoted to study the greybody factor and absobtion coefficient. We summarize our results in Sect.~\ref{conclusion}. Throughout the paper we use geometrized units where $G=c=1$ and a metric with signature $(-,+,+,+)$.

\section{Non-linear magnetic-charged black hole surrounded by quintessence \label{bhspacetime}}

A non-rotating magnetic-charged black hole surrounded with quintessence was proposed in~\cite{Nam18}. The metric for this black hole solution with mass $ M $ and magnetic charge $ Q $ is given by
\begin{eqnarray}\label{mainmetric}
ds^{2}=-f(r)dt^{2}+f(r)^{-1}dr^{2}+r^{2}d\Omega^{2} \\
f(r)=1-\frac{2Mr^{2}}{r^{3}+Q^{3}}-\frac{c}{r^{3\omega_{q}+1}}\ .
\end{eqnarray}
Here $ \omega_{q} (-1<\omega_{q}<-1/3) $ is the quintessential state parameter and $ c $ is a positive normalisation constant. In the case $ c=0 $, we can obtain the non-linear magnetic-charged black hole in the flat background or the Hayward-like black hole ~\cite{Hayward06}. 

We can see the behaviour of the black hole solution at large and short distances. For large distances, i.e. when $ Q<<r $, the solution corresponds to a weak non-linear magnetic field. The metric coefficient $ f(r) $ becomes
\begin{equation}
f(r)=1-\frac{2M}{r}-\frac{c}{r^{3\omega_{q}+1}}\ .
\end{equation}
Therefore at large distances, the solution behaves as a Schwarzschild black hole surrounded by quintessence ~\cite{Kiselev03}. At short distances, i.e. $ Q>>r $, it corresponds to a strong non-linear magnetic field. The metric coefficient become
\begin{equation}\label{desitt}
f(r)=1-\frac{\Lambda}{3}r^{2}-\frac{c}{r^{3\omega_{q}+1}},
\end{equation} 
where $ \Lambda=6M/Q^{3}>0 $. Equation~\eqref{desitt} behaves like dS geometry, the pressure is negative and thus we can avoid a singular end-state of the gravitationally collapsed matter. On the other hand, the curvature singularity of the gravitationally collapsed matter should disappear and be replaced by a dS-like geometry core.

\begin{figure*}[t]
	\begin{center}
		\includegraphics[type=pdf,ext=.pdf,read=.pdf,width=5.5cm]{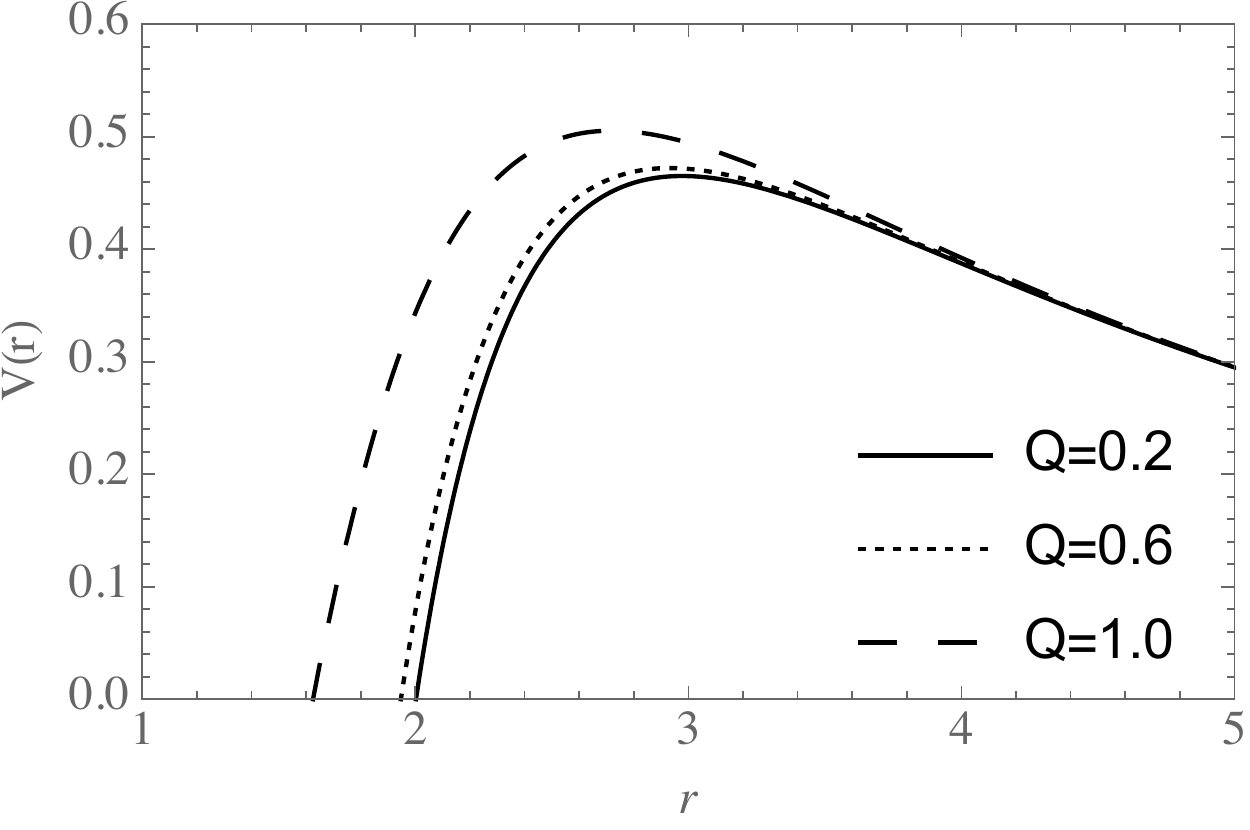}
		\includegraphics[type=pdf,ext=.pdf,read=.pdf,width=5.5cm]{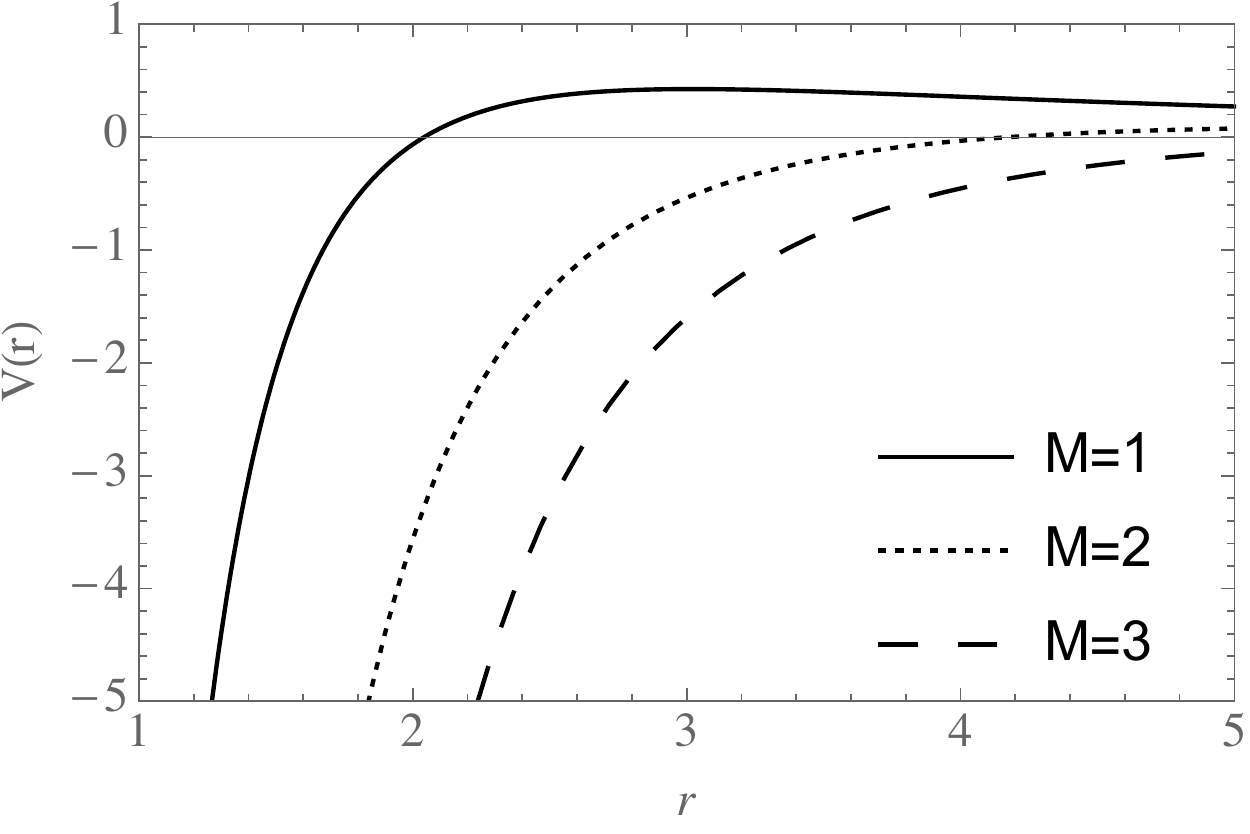}
		\includegraphics[type=pdf,ext=.pdf,read=.pdf,width=5.5cm]{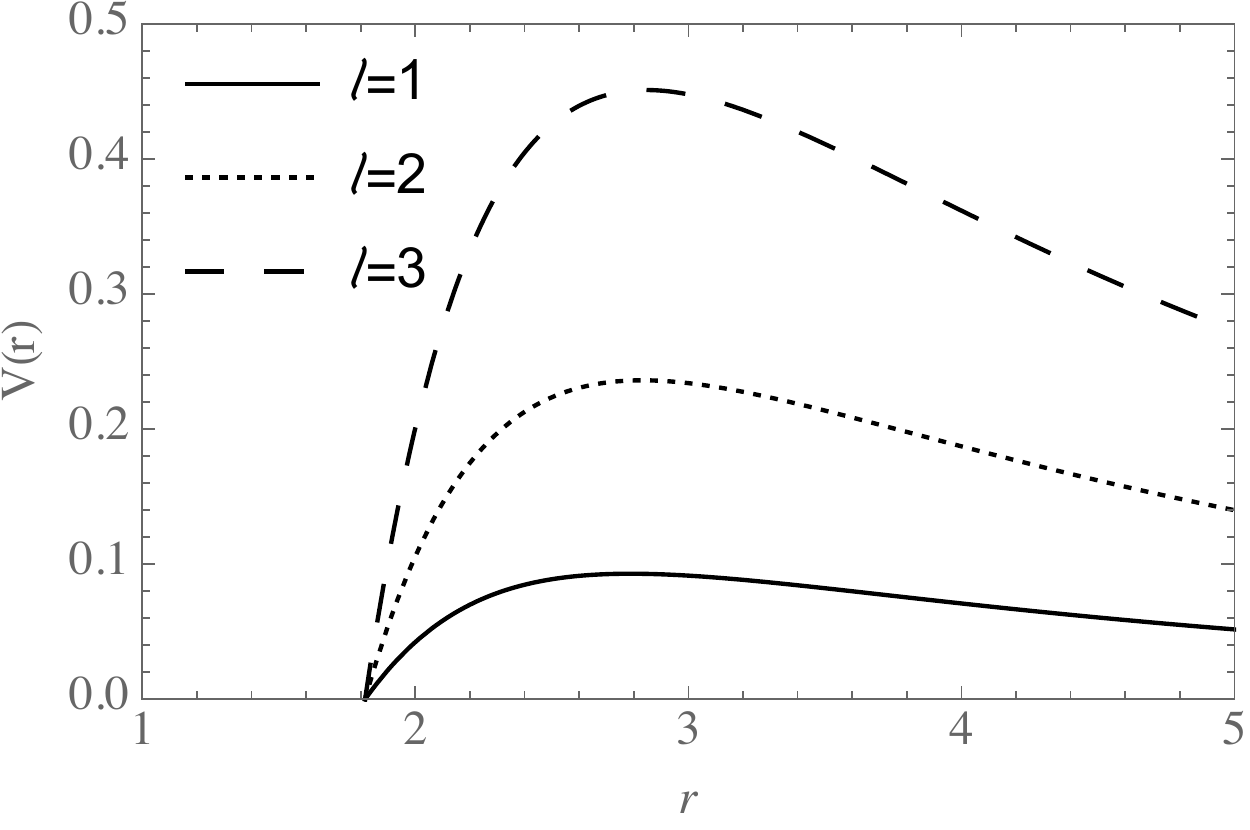}
	\end{center}
	\vspace{-0.5cm}
	\caption{Left panel: Behavior of the effective potential $ V(r) $ with the charge. Here $ l = 3 $, $ M = 1 $ and $ c = 0.001 $. The solid line, the dotted line and the dashed line correspond to $ Q = 0.2 $, $ Q = 0.6 $ and $ Q = 1.0 $ respectively. Mid panel: Behavior of the effective potential $ V(r) $ with mass. Here $ l = 3 $, $ Q = 0.1 $ and $ c = 0.01 $. The solid line, the dotted line and the dashed line corresponds to $ M = 1 $, $ M = 2 $ and $ M = 3 $ respectively. Right panel: Behavior of the effective potential $ V(r) $ with the spherical harmonic index. Here $ M = 1 $, $ Q = 0.9 $ and $ c = 0.01 $. The solid line, the dotted line and the dashed line corresponds to $ l = 1 $, $ l = 2 $ and $ l = 3 $ respectively.  \label{pots} }
\end{figure*}

\section{Massless Scalar perturbation\label{scalarpert}}

In this section, we shall briefly write the equations of scalar perturbations of a non-linear magnetic-charged black hole surrounded by quintessence~\eqref{mainmetric}. It was pointed out in~\cite{Konoplya11} that, if there is no backreaction on the background, the perturbations of black hole spacetimes can be studied not only by adding the perturbation terms into the spacetime metric, but also by introducing fields to the spacetime metric. 

The equation of motion of a massless scalar field $ \Phi $ in curved spacetime is given by the Klein-Gordon equation,

\begin{equation}\label{kgeqn}
\frac{1}{\sqrt{-g}}\partial_{\mu}( \sqrt{-g}g^{\mu\nu}\partial_{\nu}\Phi ) = 0.
\end{equation}

Here $ \Phi $ is a function of the coordinates $ ( t, r, \theta, \phi ) $. Since the metric is spherically symmetric, the field evolution should be independent of rotations. In order to separate the variables, we write the wave function as

\begin{equation}
\Phi = Y(\theta, \phi)\frac{\psi(r,t)}{r},
\end{equation}
where $ Y(\theta,\phi) $ are the spherical harmonics. Now,~\eqref{kgeqn} simplifies to 

\begin{equation}\label{mastereq}
-\frac{\partial^{2}\psi}{\partial t^{2}} + \frac{\partial^{2}\psi}{\partial r^{2}_{*}}+\Big(\omega^{2}-V(r)\Big)\psi (r) = 0,
\end{equation}
 where,
 
 \begin{equation}\label{poten}
 V(r) = \frac{l(l+1)f(r)}{r^{2}} + \frac{f(r)f'(r)}{r}.
 \end{equation}

Here $ l $ is the spherical harmonic index and  $ r_{*} $ is the well-known tortoise coordinate, given by

\begin{equation}
dr_{*}=\frac{dr}{f(r)}.
\end{equation}

Here, $ r_{*} $ cannot be evaluated explicitly, because of the nature of the function $ f(r) $. When $ r \rightarrow r_c $, $ r_{*} \rightarrow \infty $. When $ r \rightarrow r_{+} $, $ r_{*} \rightarrow -\infty $. Here $ r_{+} $ is the event horizon and $ r_c $ is the cosmological horizon.

The effective potential $ V(r) $ is plotted in Fig.~\ref{pots} to see how it changes with the charge $ Q $, the mass $ M $ and the spherical harmonic index $ l $. The first panel shows the behaviour of the potential for different values of the charge parameter. We can see that the height of the potential increases with the magnetic charge and this represents the suppression of emission modes for higher charge values. The middle panel shows the change in the potential when mass is varied and we can see that the potential decreases for an increase in mass. The right panel shows the increase of the peak of potential when we increase the spherical harmonic index. This again represents the suppression of scalar emission modes for higher values of $ l $.

\section{Numerical Results\label{numresult}}

\subsection{Evolution of scalar perturbations}

In order to study the scalar perturbations of a non-linear magnetic charged black hole surrounded by quintessence, we rewrite the wave equation for the propagation of scalar perturbations (\ref{mastereq}) in null coordinates defined by,

\begin{equation}
du=dt-dr_* \ , \ \ \ \ dv=dt+dr_*.
\end{equation}

The wave equation becomes,

\begin{equation}\label{nullmaster}
\Bigg(4 \frac{\partial^{2}}{\partial u \partial v} + V(u,v)\Bigg) \psi(u,v) =0\ .
\end{equation}

We solve equation (\ref{nullmaster}) numerically with a simple initial condition, i.e., a Gaussian pulse profile centered around $ v_{c} $ and width $ \sigma $,

\begin{equation}
\psi(u=u_{0},v) = \exp\Bigg[ - \frac{(v-v_{c})^{2}}{2\sigma^{2}} \Bigg]\ .
\end{equation}

The integration is done on a triangular grid. This method is extensively reviewed in \cite{Abdalla06b,Toshmatov17a,Chirenti07}. We follow~\cite{Chirenti07} for discretization of the wave function. During the integration of equation~(\ref{nullmaster}), we extract the values of the field $ \psi $ at constant $ r_{*} $ and allow the field to evolve to larger values of $ t $.    

\begin{figure*}[t]
	\begin{center}
		\includegraphics[type=pdf,ext=.pdf,read=.pdf,width=5.5cm]{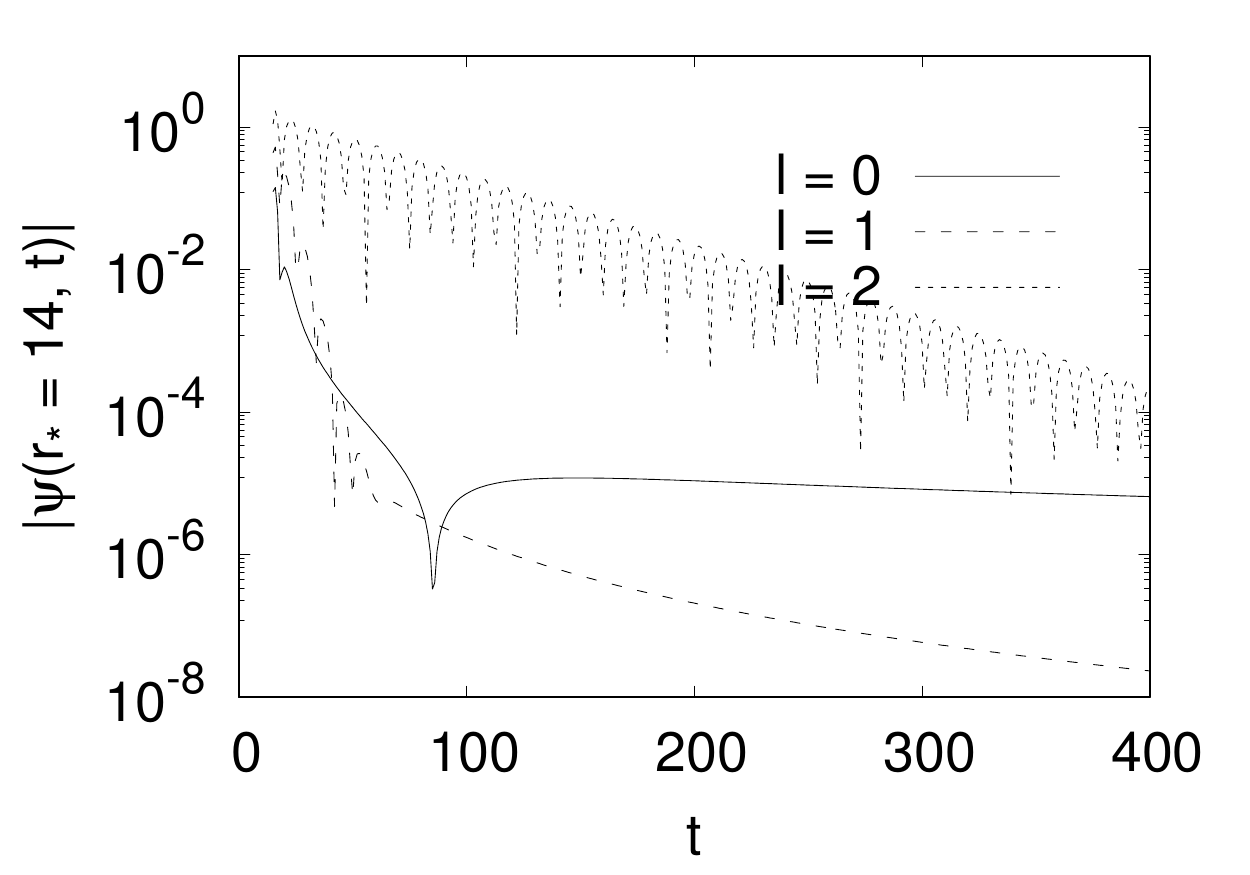}
		\includegraphics[type=pdf,ext=.pdf,read=.pdf,width=5.5cm]{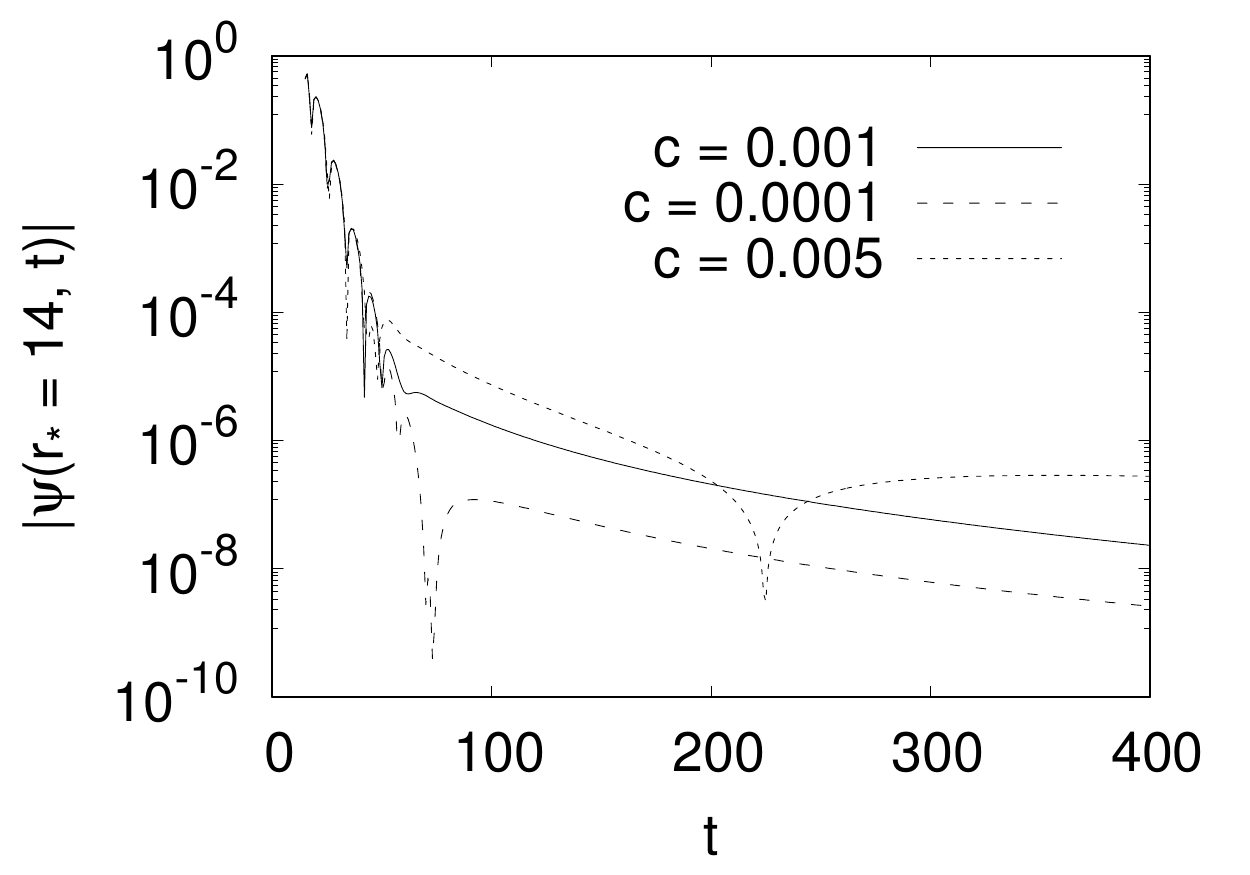}
		\includegraphics[type=pdf,ext=.pdf,read=.pdf,width=5.5cm]{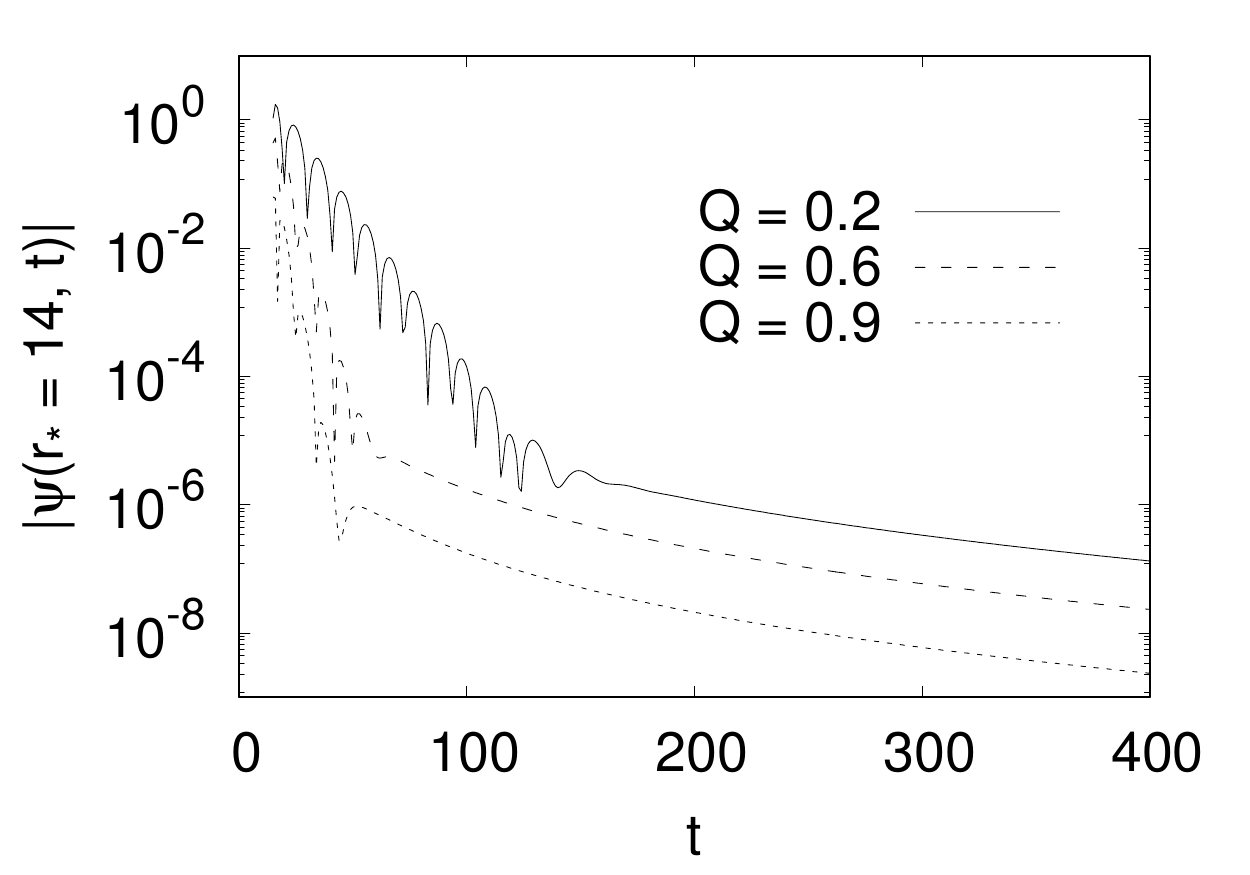}
	\end{center}
	\vspace{-0.5cm}
	\caption{Left panel: Time evolution behavior of scalar perturbations for $ l = 0 $, $ l = 1 $ and $ l = 2 $. Mid panel: Time evolution behavior of scalar perturbations for $ c = 0.001 $, $ c = 0.0001 $, $ c = 0.005  $. Right panel: Time evolution behavior of scalar perturbations for $ Q = 0.2 $, $ Q = 0.6 $, $ Q = 0.9 $. \label{pertev} }
\end{figure*}

We plot the time evolution of scalar perturbations in Fig.~\ref{pertev} for different parameters. The left panel shows that scalar perturbations with larger $ l $ live longer. In the middle panel, we plot $ \psi $ for different values of $ c $ and we see that the perturbations start to become unstable for higher values of $ c $. In the right panel of Fig.~\ref{pertev}, we can see the behaviour of the perturbations with respect to the  charge~$Q$.

\subsection{Quasi-normal modes}

One of our main goals in this paper is to study the quasi-normal modes (QNM) and the stability of the perturbations in a non-linear magnetic-charged black hole spacetime surrounded by quintessence. We shall focus on massless scalar perturbations here.

QNMs for a perturbed black hole space-time are the solutions to the wave equation given in~\eqref{mastereq}. In order to obtain these solutions, one has to impose proper boundary conditions. We
impose that the wave at the horizon is purely incoming and that the wave at spatial infinity is purely outgoing:
\begin{equation}
	\begin{aligned}
		& \psi (r) \sim e^{-i\omega r_{*}} \ \ \   \text{as} \ \ \  r_{*} \rightarrow - \infty (r\rightarrow r_{h}) ,\\
		& \psi (r) \sim e^{i\omega r_{*}} \ \ \   \text{as} \ \ \  r_{*} \rightarrow  + \infty (r\rightarrow r_{c}) ,\\
	\end{aligned}
\end{equation}
where $ r_{h} $ and $ r_{c} $ are event horizon and cosmological horizon, respectively. 

It is impossible to analytically solve the time-independent, second order differential equation \eqref{mastereq} with the potential~\eqref{poten} for a non-linear magnetic-charged black hole surrounded by quintessence. Therefore, we use the sixth order WKB method for numerical
calculations which is given by the relation

\begin{equation}
\frac{i(\omega^{2}-V_{0})}{\sqrt{-2V''_{0}}}+ \sum\limits_{j=2}^{6} \Lambda_{j} = n + \frac{1}{2},
\end{equation}
where a prime denotes derivative with respect to the tortoise coordinate $ r_{*} $ and $ V_{0} $ stands for the value of effective potential at its local maxima. $ j $ denotes the order of WKB approximation and $ \Lambda_{j} $ is a correction term corresponding to the $ j $th order. Expressions for $ \Lambda_{j} $ can be found in~\cite{Iyer87,Konoplya03}.

\begin{figure*}[t]
	\begin{center}
		\includegraphics[type=pdf,ext=.pdf,read=.pdf,width=7.0cm]{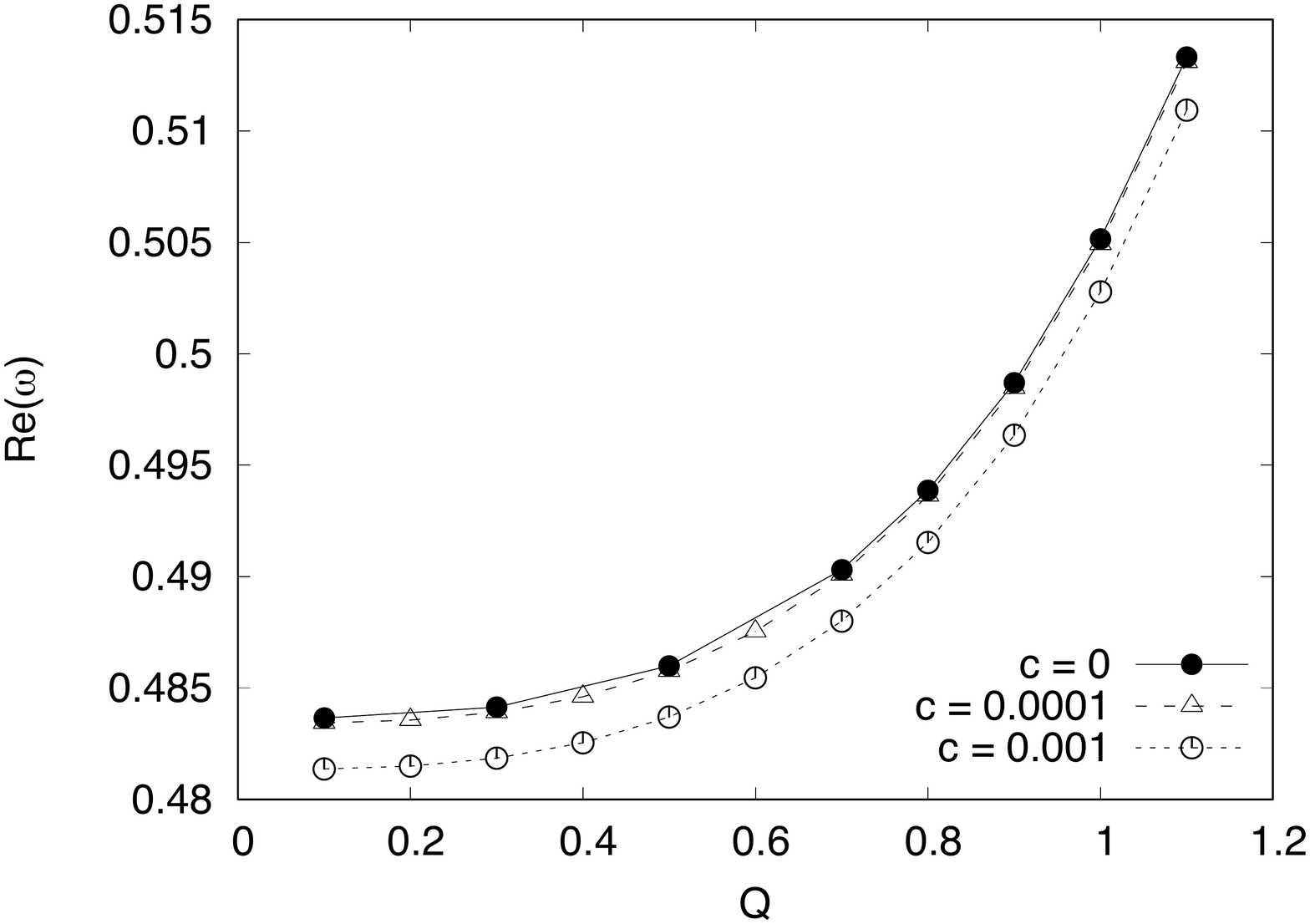}
		\includegraphics[type=pdf,ext=.pdf,read=.pdf,width=7.0cm]{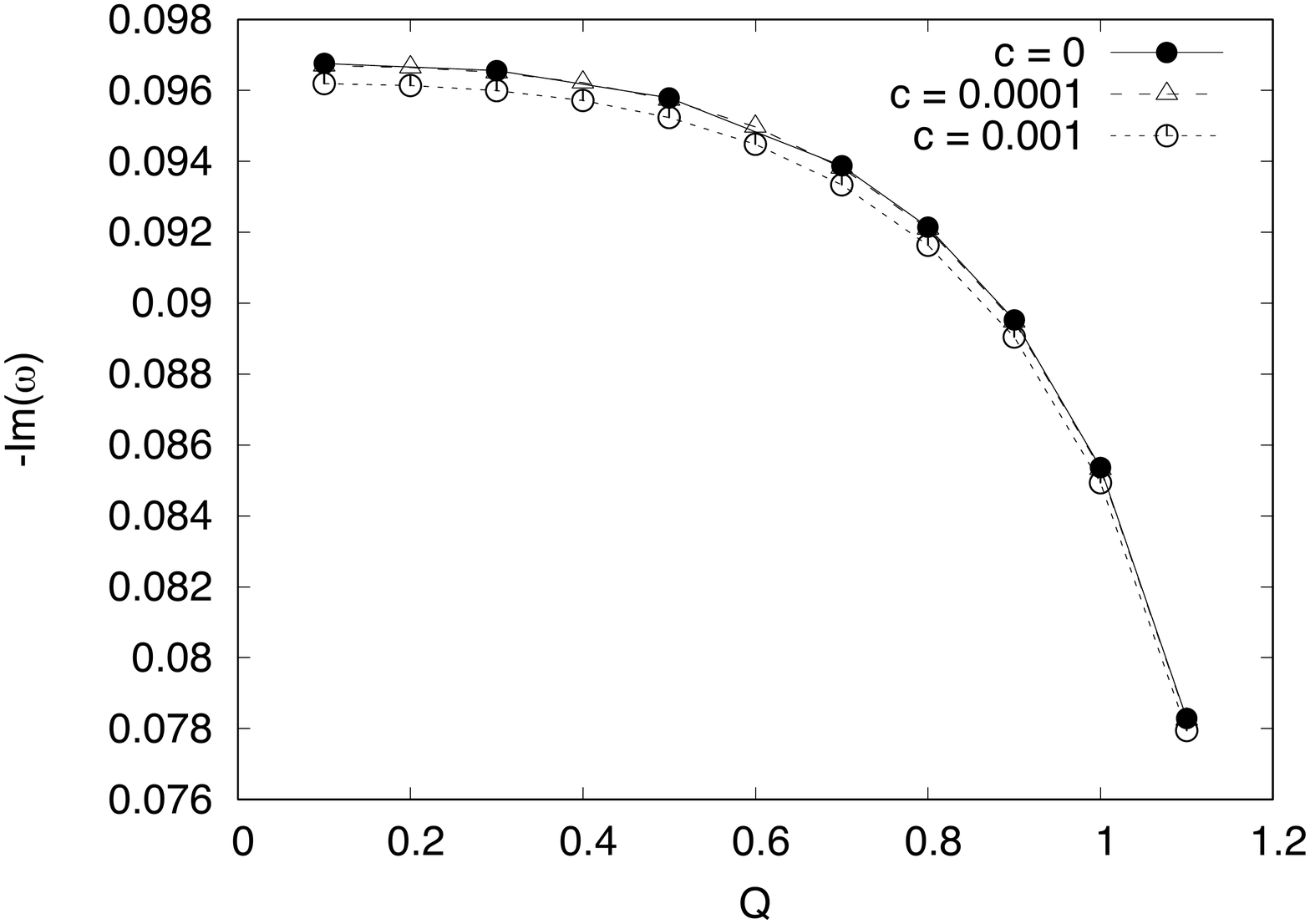}
	\end{center}
	\vspace{-0.5cm}
	\caption{The left panel shows the dependence of real part of quasi-normal frequencies on $ Q $ for $ c=0 $, $ c=0.0001 $ and $ c=0.001 $. The right panel shows the dependence of imaginary part of quasi-normal frequencies on $ Q $ for $ c=0 $, $ c=0.0001 $ and $ c=0.001 $.  \label{qnm_charge} }
\end{figure*}

\begin{figure*}[t]
	\begin{center}
		\includegraphics[type=pdf,ext=.pdf,read=.pdf,width=7.0cm]{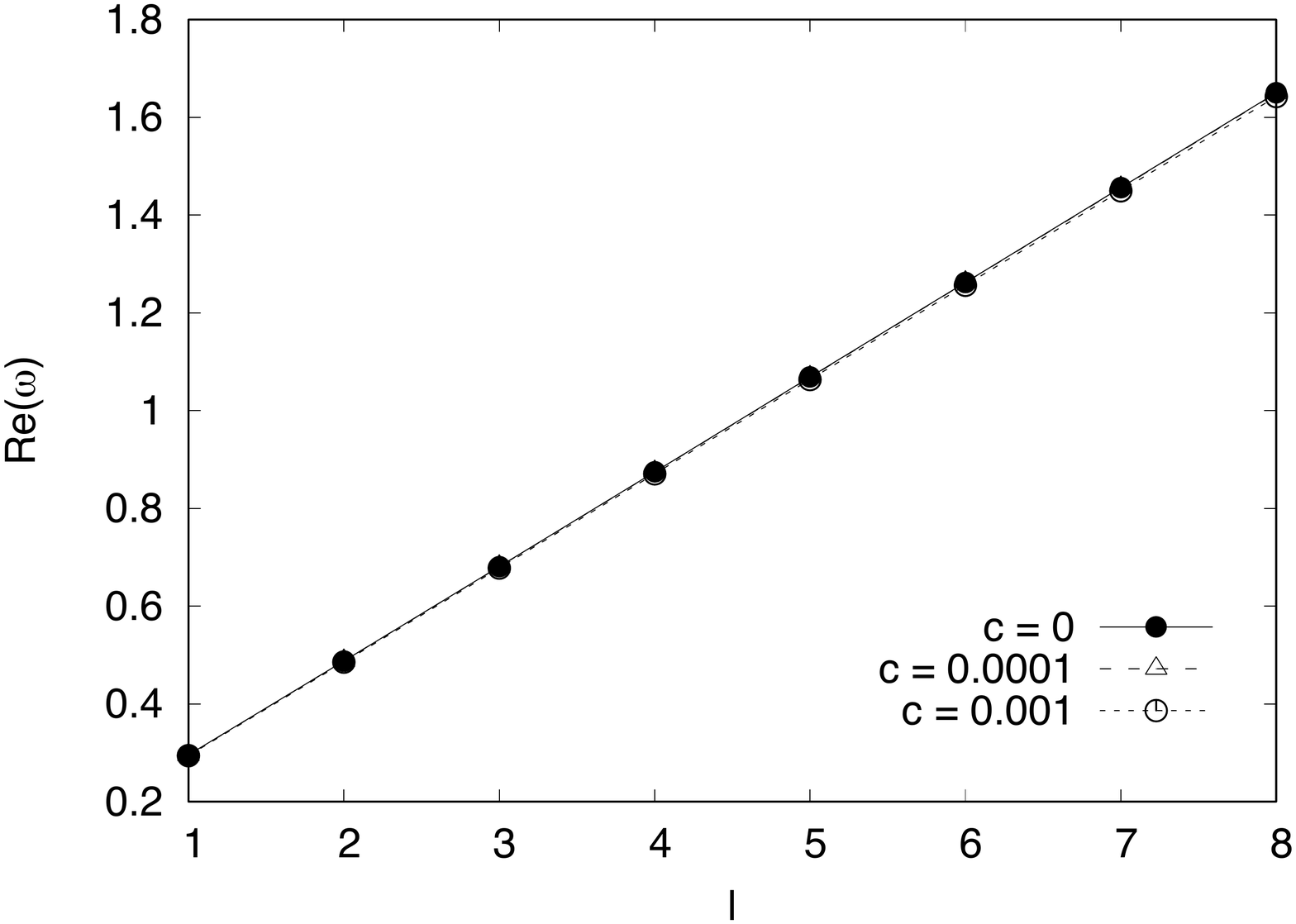}
		\includegraphics[type=pdf,ext=.pdf,read=.pdf,width=7.0cm]{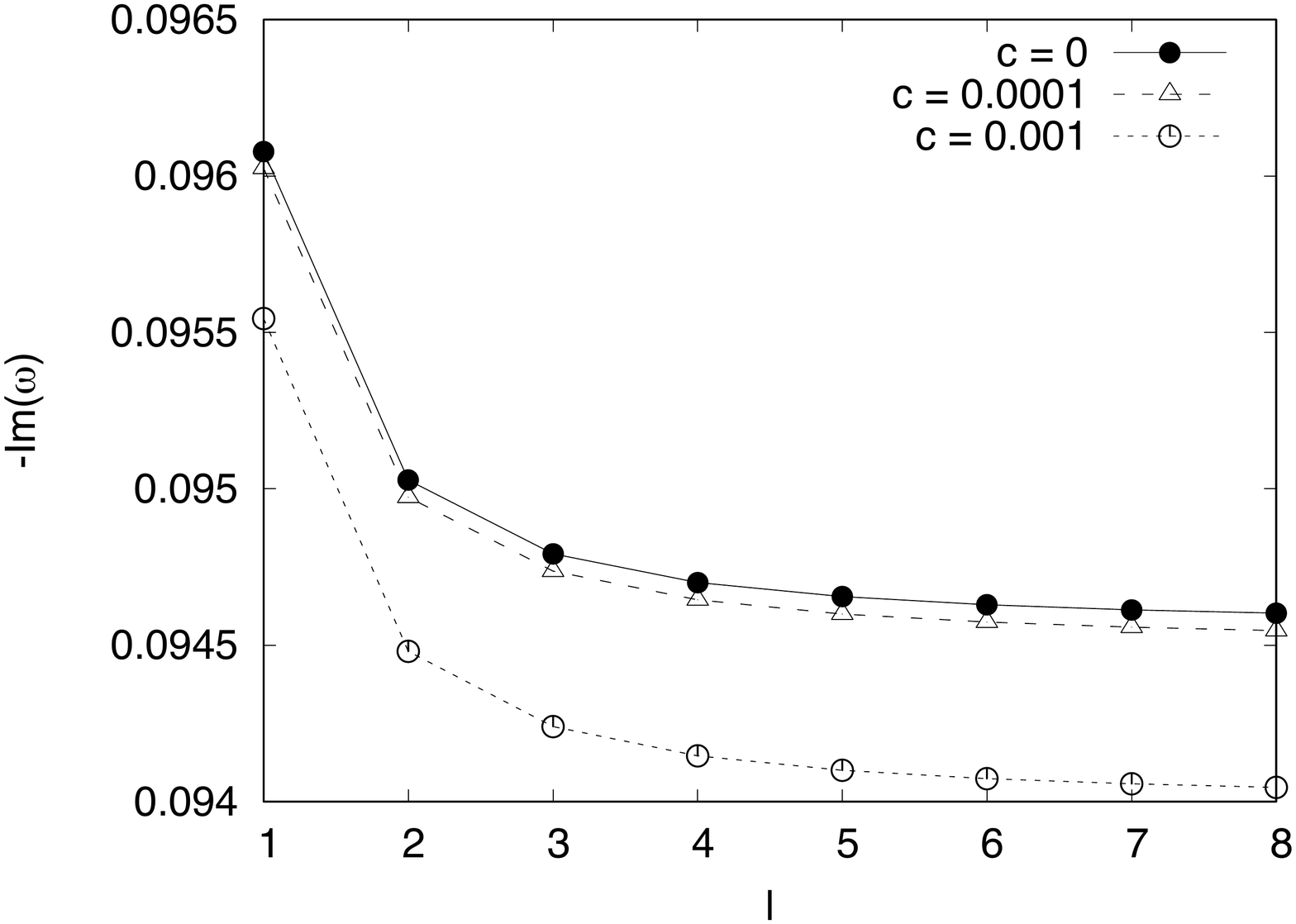}
	\end{center}
	\vspace{-0.5cm}
	\caption{The left panel shows the dependence of real part of quasi-normal frequencies on $ l $ for $ c=0 $, $ c=0.0001 $ and $ c=0.001 $. The right panel shows the dependence of imaginary part of quasi-normal frequencies on $ l $ for $ c=0 $, $ c=0.0001 $ and $ c=0.001 $.  \label{qnm_l} }
\end{figure*}

\begin{figure*}[t]
	\begin{center}
		\includegraphics[type=pdf,ext=.pdf,read=.pdf,width=7.0cm]{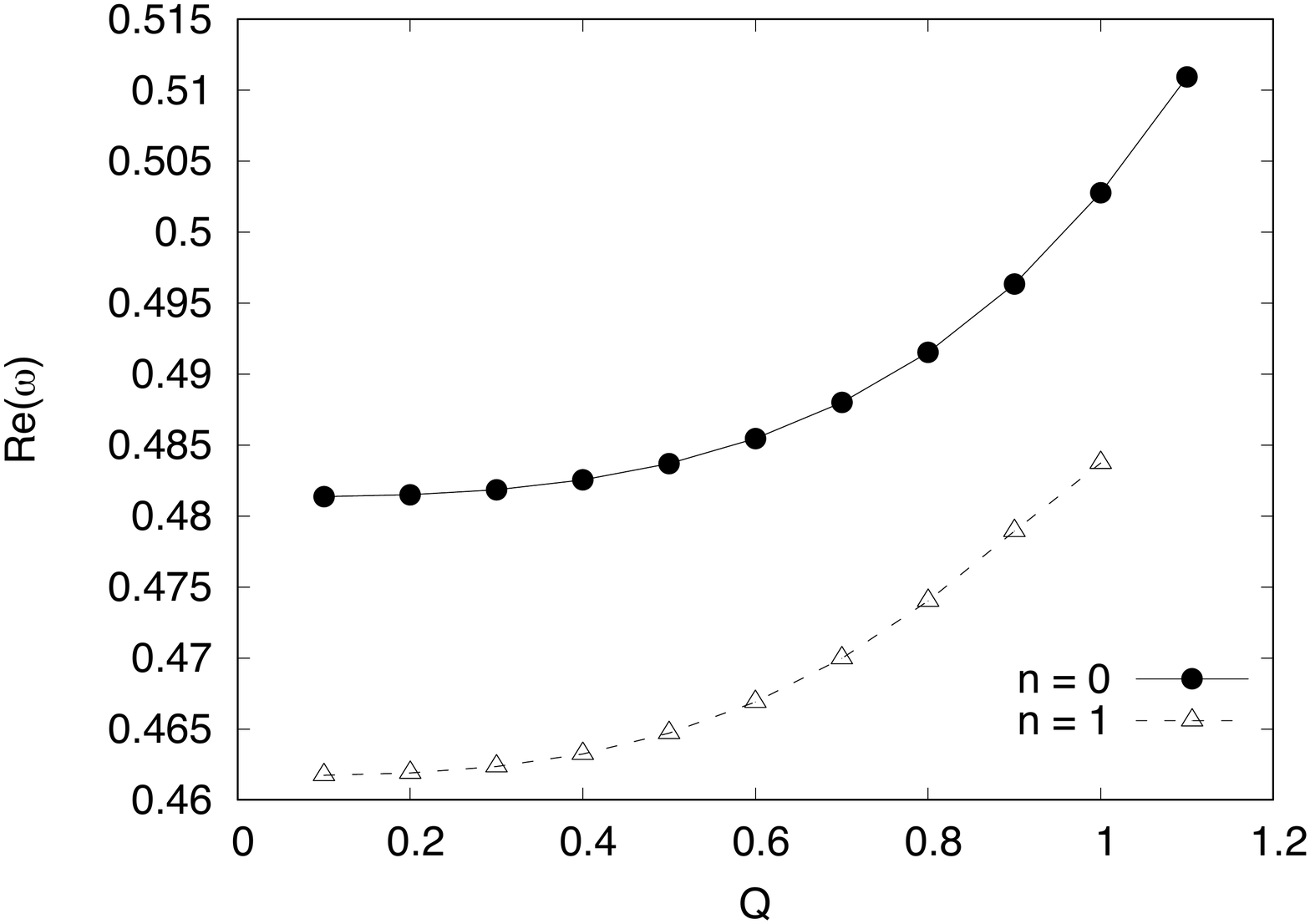}
		\includegraphics[type=pdf,ext=.pdf,read=.pdf,width=7.0cm]{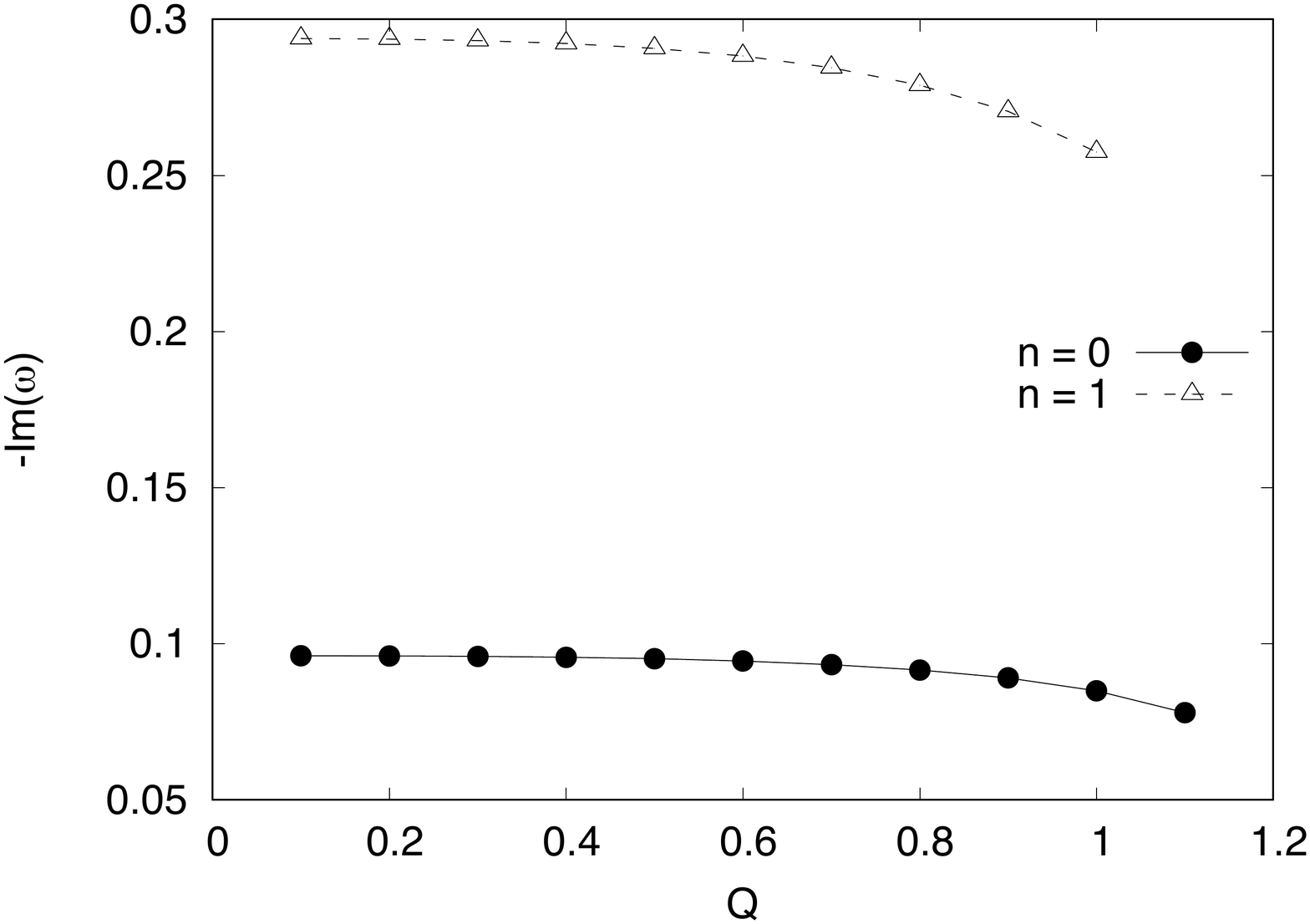}
	\end{center}
	\vspace{-0.5cm}
	\caption{ The left panel shows the dependence of real part of quasi-normal frequencies on $ Q $ for $ n = 0 $ and $ n = 1 $. The right panel shows the dependence of imaginary part of quasi-normal frequencies on $ Q $ for $ n = 0 $ and $ n = 1 $.    \label{qnm_n} }
\end{figure*}

\begin{table}
	\begin{center}
	    \begin{tabular}{ c c c c } 
		\hline
		Q \ \ & \ \ $c$ \ \ & \ \ Re$(\omega)$ \ \ & \ \ Im$(\omega)$ \ \ \\ 
		\hline
		0.1 \ \ & \ \ 0.0 \ \  & \ \ 0.48366 \ \ & \ \ -0.096758 \\ 
		
		0.1 \ \ & \ \ 0.0001 \ \ & \ \ 0.483432 \ \ & \ \ -0.096701 \\
		
		0.1  \ \ & \ \ 0.001 \ \ & \ \ 0.48137 \ \ & \ \ -0.096191 \\
		
		0.5  \ \ &  \ \ 0.0 \ \ & \ \ 0.485999 \ \ & \ \ -0.095791 \\
		
		0.5 \ \ & \ \ 0.0001 \ \ & \ \ 0.485769 \ \ & \ \ -0.095736 \\ 
		
		0.5 \ \ & \ \ 0.001 \ \ & \ \ 0.483698 \ \ & \ \ -0.095235 \\
		
		0.9 \ \ & \ \ 0.0 \ \ & \ \ 0.498703 \ \ & \ \ -0.089525 \\
		
		0.9 \ \  & \ \ 0.0001 \ \ & \ \ 0.498468 \ \ & \ \ -0.089477 \\
		
		0.9 \ \ & \ \ 0.001 \ \ & \ \ 0.496348 \ \ & \ \ -0.089045 \\ 
		\hline
		\end{tabular}
		\caption{Fundamental quasinormal frequencies ($n = 0$) for scalar perturbations of a non-linear magnetic charged black hole surrounded by quintessence. Here $ \omega_{q} = -2/3 $, $ l = 2 $.}
		\label{tab1}
	\end{center}
\end{table}

\begin{table}
	\begin{center}
		\begin{tabular}{ c c c c } 
			\hline
			$ l $ \ \ & \ \ $c$ \ \ & \ \ Re$(\omega)$ \ \ & \ \ Im$(\omega)$ \ \ \\ 
			\hline
			1 \ \ & \ \ 0.0 \ \  & \ \ 0.295528 \ \ & \ \ -0.096078 \\ 
			
			1 \ \ & \ \ 0.0001 \ \ & \ \ 0.295377 \ \ & \ \ -0.096024 \\
			
			1  \ \ & \ \ 0.001 \ \ & \ \ 0.294019 \ \ & \ \ -0.095544 \\
			
			2  \ \ &  \ \ 0.0 \ \ & \ \ 0.487769 \ \ & \ \ -0.095027 \\
			
			2 \ \ & \ \ 0.0001 \ \ & \ \ 0.487538 \ \ & \ \ -0.094973 \\ 
			
			2 \ \ & \ \ 0.001 \ \ & \ \ 0.485461 \ \ & \ \ -0.094480 \\
			
			3 \ \ & \ \ 0.0 \ \ & \ \ 0.681045 \ \ & \ \ -0.094792 \\
			
			3 \ \  & \ \ 0.0001 \ \ & \ \ 0.68073 \ \ & \ \ -0.094737 \\
			
			3 \ \ & \ \ 0.001 \ \ & \ \ 0.677896 \ \ & \ \ -0.094240 \\ 
			\hline
		\end{tabular}
		\caption{Fundamental quasinormal frequencies ($n = 0$) for scalar perturbations of a non-linear magnetic charged black hole surrounded by quintessence. Here $ \omega_{q} = -2/3 $, $ Q = 0.6 $.}
		\label{tab2}
	\end{center}
\end{table}

In Table~\ref{tab1} and Table~\ref{tab2}, we list the fundamental quasinormal frequencies $ n = 0 $ of scalar perturbations of a non-linear magnetic charged black hole surrounded by quintessence. Table~\ref{tab1} shows the variation with respect to the charge for different values of $ c $.  The real part of the frequency decreases with increasing $ c $. With increasing charge the real part of the frequency increases but the magnitude of the imaginary part of the frequency decreases. In Table~\ref{tab2}, we list the fundamental quasinormal frequencies with respect to smaller spherical harmonic index $ l $ for different values of $ c $. We can see that the real part of the frequency decreases and the magnitude of the imaginary part of the frequency also decreases with increase in $ c $ for the same $ l $. Here we have to note that the WKB method has low accuracy in the small $ l $ regime.

In Fig.~\ref{qnm_charge} we plot the quasi-normal frequencies of scalar perturbations with respect to the charge parameter $ Q $ for $ c=0 $, $ c=0.0001 $ and $ c=0.001 $. The left panel shows the real part of the frequency and the right panel shows the imaginary part of the frequency. Here we consider $ M=1 $ and $ l=2 $  and $ \omega_{q}=-2/3 $.

\begin{figure*}[t]
	\begin{center}
		\includegraphics[type=pdf,ext=.pdf,read=.pdf,width=7.0cm]{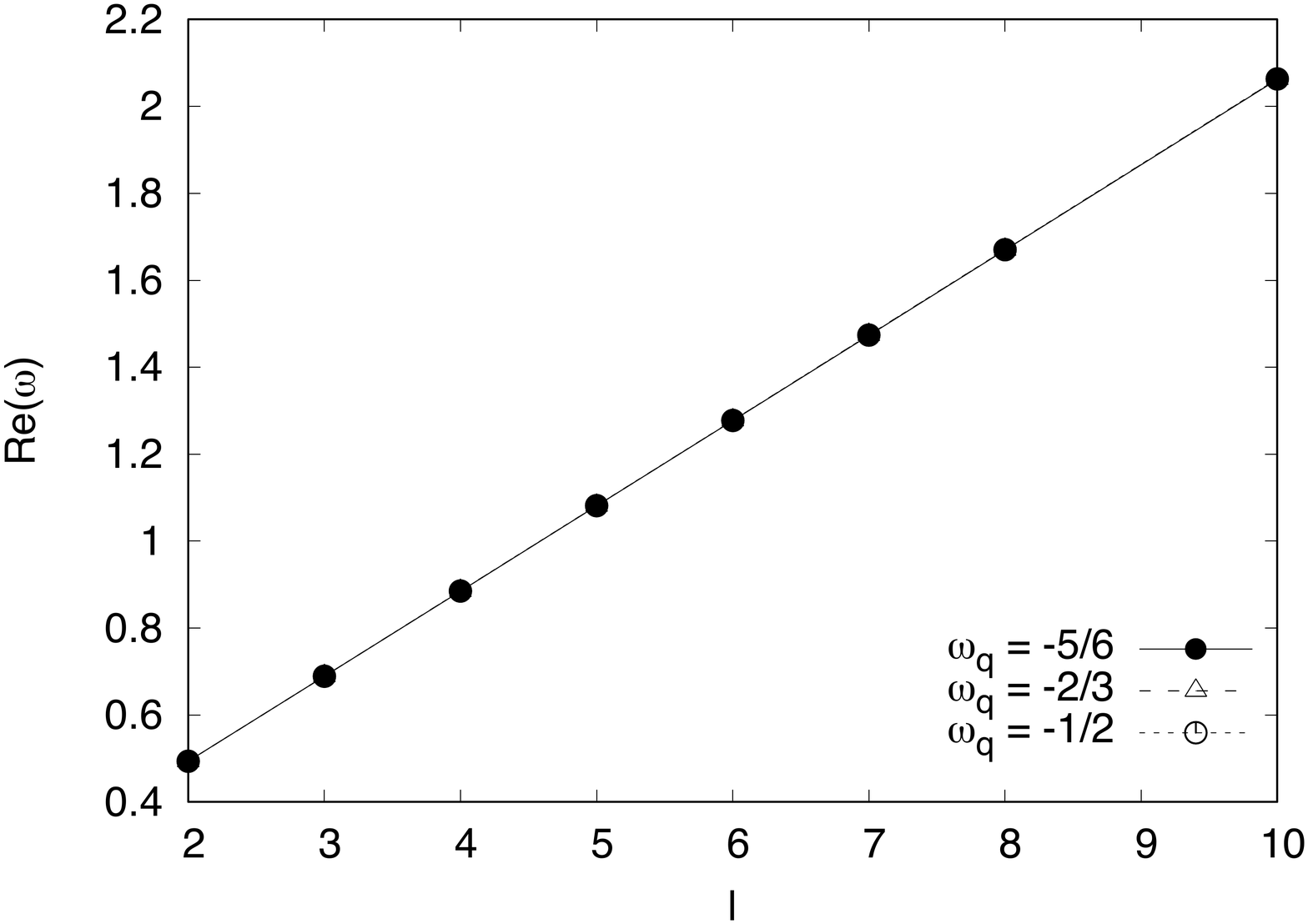}
		\includegraphics[type=pdf,ext=.pdf,read=.pdf,width=7.0cm]{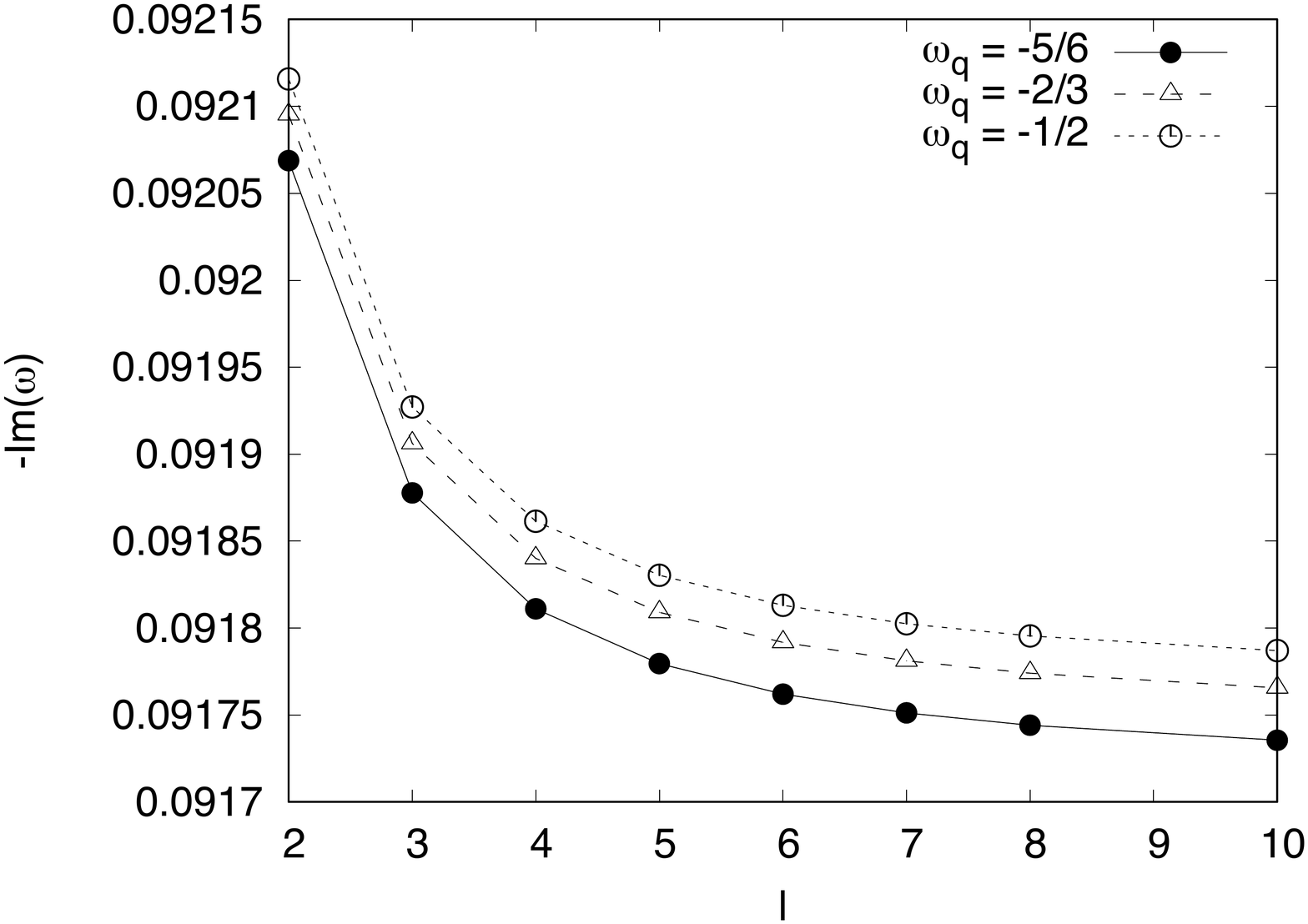}
	\end{center}
	\vspace{-0.5cm}
	\caption{ The left panel shows the dependence of real part of quasi-normal frequencies on $ l $ for $ \omega_{q} = -5/6 $, $ \omega_{q} = -2/3 $ and $ \omega_{q} = -1/2 $. The right panel shows the dependence of imaginary part of quasi-normal frequencies on $ l $ for $ \omega_{q} = -5/6 $, $ \omega_{q} = -2/3 $ and $ \omega_{q} = -1/2 $. Here $ M = 1 $, $ n = 0 $, $ Q = 0.8 $ and $ c = 0.0001 $.     \label{qnm_w} }
\end{figure*}

\begin{figure*}[t]
	\begin{center}
		\includegraphics[type=pdf,ext=.pdf,read=.pdf,width=7.0cm]{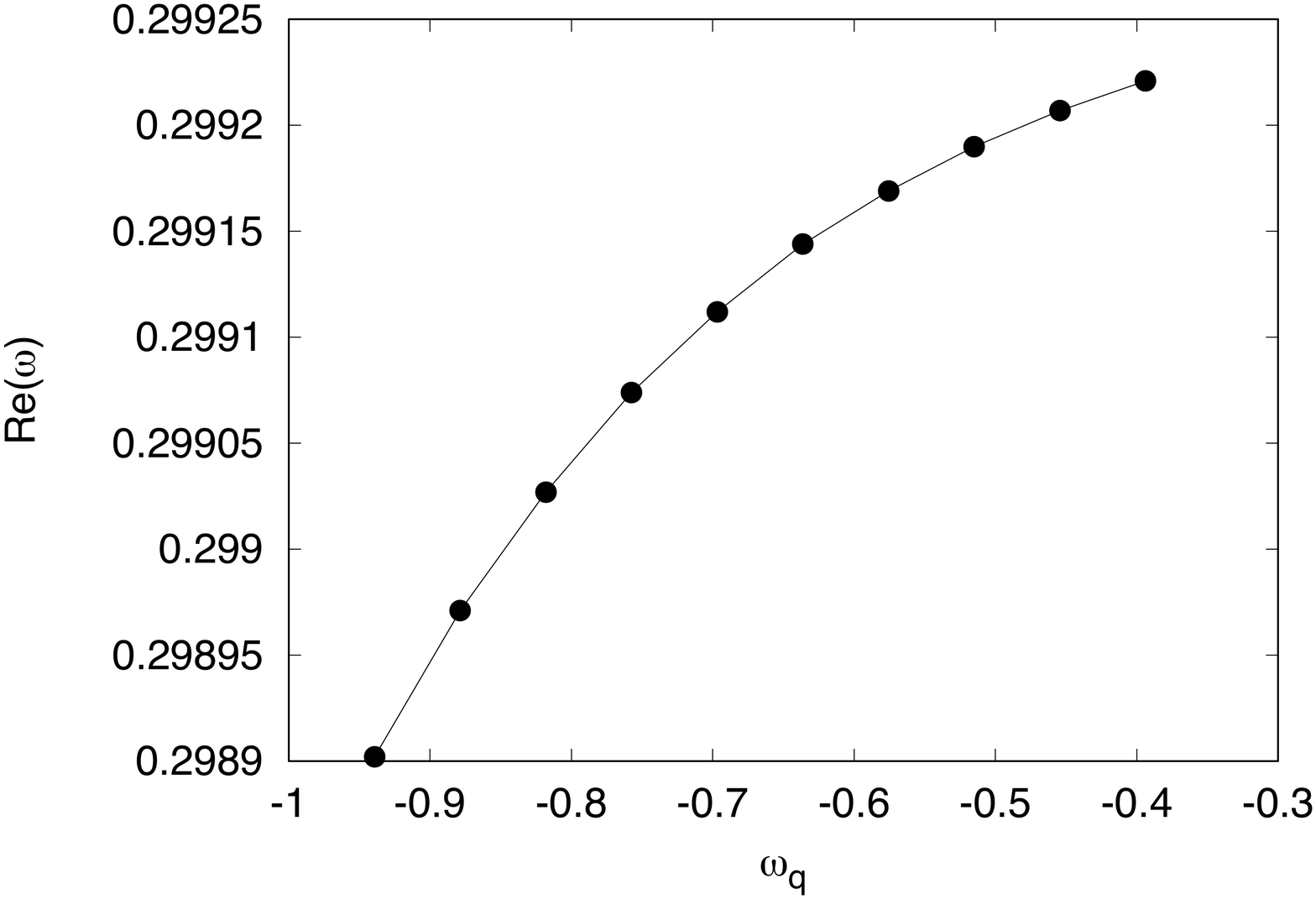}
		\includegraphics[type=pdf,ext=.pdf,read=.pdf,width=7.0cm]{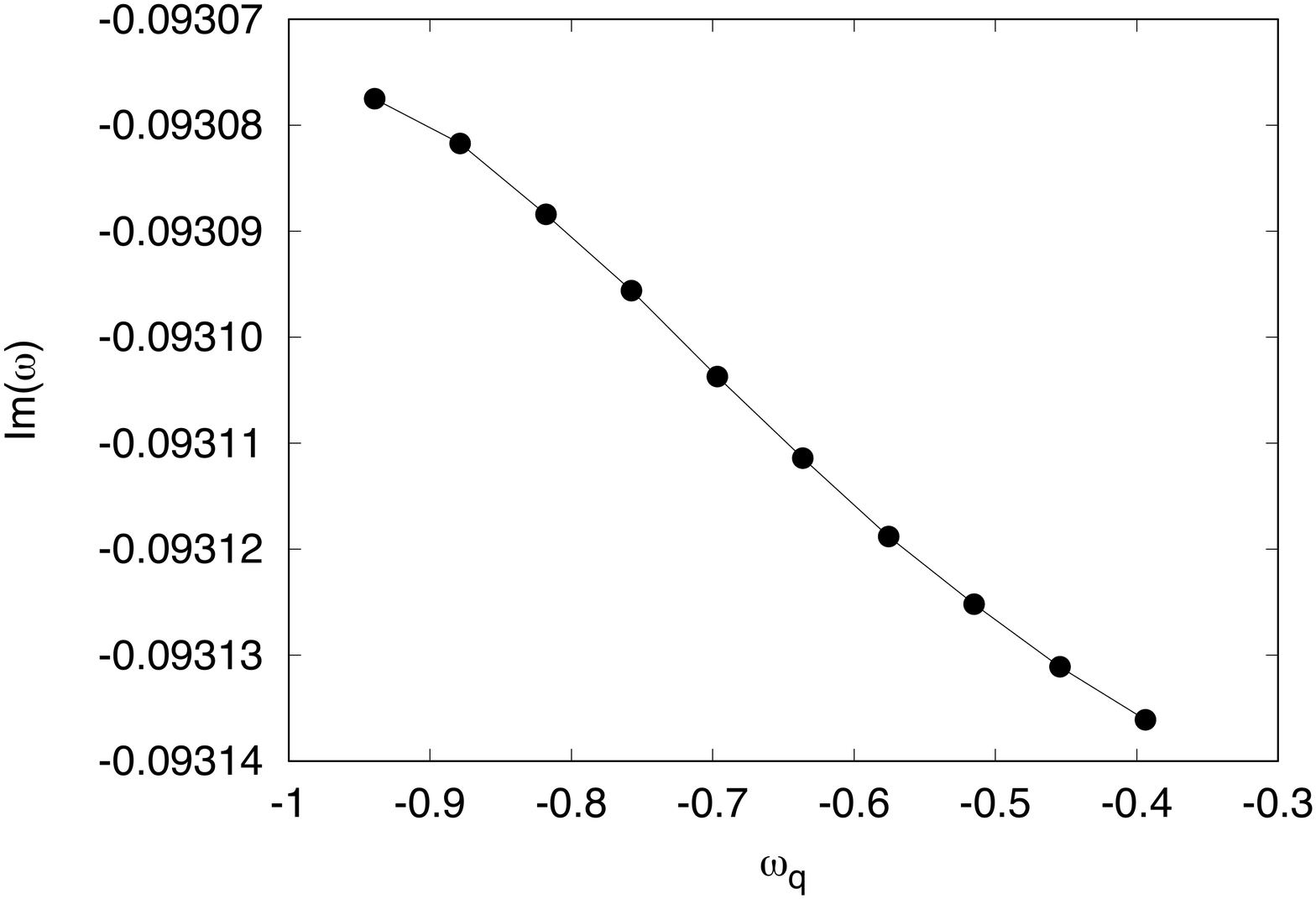}
	\end{center}
	\vspace{-0.5cm}
	\caption{ The left panel shows the dependence of real part of quasi-normal frequencies on $ \omega_{q} $. The right panel shows the dependence of imaginary part of quasi-normal frequencies on $ \omega_{q} $. Here $ M = 1 $, $ l = 10 $, $ n = 0 $, $ Q = 0.8 $ and $ c = 0.0001 $.    \label{qnm_w1} }
\end{figure*}

In Fig.~\ref{qnm_l}, we plot the quasi-normal frequencies with respect to spherical harmonic index $ l $ for $ c=0 $, $ c=0.0001 $ and $ c=0.001 $. The left panel shows the real frequencies and the right panel shows the imaginary frequencies. Here we consider $ M = 1 $, $ Q = 0.6 $ and $ \omega_{q}=-2/3 $. 

In Fig.~\ref{qnm_n}, we plot the quasi-normal frequencies with respect to the charge parameter $ Q $ for $ n=0 $, $ n=1 $. The left panel shows the real frequencies and the right panel shows the imaginary frequencies. Here we consider $ M = 1 $, $ c = 0.001 $, $ \omega_{q}=-2/3 $ and $ l=2 $. 

In Fig.~\ref{qnm_w}, we plot the quasi-normal frequencies with respect to spherical harmonic index for different values of quintessential parameter $ \omega_{q} $. The left panel shows the increase of real part of the frequency with $ l $ and we see that the change with respect to $ \omega_{q} $ is very small. The right panel shows the magnitude of imaginary part of quasi-normal frequencies which decreases with increasing $ l $. 

In Fig.~\ref{qnm_w1}, we plot the quasi-normal frequencies with respect to quintessential parameter $ \omega_{q} $. The left panel shows the increase of real part of the quasi-normal frequency with the increase in quintessential parameter $ \omega_{q} $ and the right panel shows the decrease of imaginary part of quasi-normal with the increase in quintessential parameter.

\section{Greybody Factors and Absorbtion coefficient\label{greybody}}

\subsection{Nature of greybody factors}

In this subsection, we shall discuss the reflection coefficients $ R(\omega) $ and transmission coefficients $ T(\omega) $ for different parameter spaces for scalar perturbations around a non-linear magnetically charged black hole surrounded by quintessence. Our analysis in this section will be analytic and we shall use a third order WKB method found in~\cite{Konoplya10a,Toshmatov16,Fernando17,Dey18} for our computations. 

Black holes are believed to be thermal systems with an associated temperature and entropy. Therefore black holes radiate and the radiation is known as Hawking radiation~\cite{Hawking75,Hawking76}. Hawking showed that, at the event horizon, the emission rate of a black hole in a mode with frequency $ \omega $ is given by

\begin{equation}
\Gamma (\omega ) = \frac{1}{e^{\beta\omega} \pm 1 }\frac{d^{3}k}{(2\pi)^{3}}.
\end{equation} 

Here, $ \beta $ is the inverse of Hawking temperature and the plus (minus) sign corresponds to fermions (bosons). But the geometry outside the event horizon might have an important effect on the emission rate measured by an observer located far away. In other words, the geometry outside the horizon will act as a potential barrier for the Hawking radiation emitted from the black hole. A part of the radiation will be tunneled through the potential barrier and reach the distant observer and the other part will be reflected back towards the black hole. The radiation recorded by the distant observer will no longer appear as a  blackbody. Mathematically, we can write emission rate measured by an observer at infinity for a frequency mode $ \omega $ as,

\begin{equation}
\Gamma (\omega ) = \frac{\gamma(\omega)}{e^{\beta\omega} \pm 1 }\frac{d^{3}k}{(2\pi)^{3}},
\end{equation}

where $ \gamma(\omega) $ is called the greybody factor. This quantity gets its name from the fact that it modifies the emitted spectrum of a black hole to a greybody. The greybody factor is naturally defined as

\begin{equation}
\gamma(\omega)=|T(\omega)|^{2}. 
\end{equation}

Now, the reflected and the transmitted waves can be represented as following

\begin{equation}
	\begin{aligned}
		& \psi(r_{*})=T(\omega) e^{-i\omega r_{*}}    \ \  ,\ \ r_{*} \rightarrow -\infty (r\rightarrow r_{h}) \\
		& \psi(r_{*})=e^{-i\omega r_{*}} + R(\omega) e^{-i\omega r_{*}}    \ \ , \ r_{*} \rightarrow +\infty (r\rightarrow r_{c}),
	\end{aligned}
\end{equation}
where $ R(\omega) $ and $ T(\omega) $ are the reflection and transmission coefficient, respectively, and they are related by
\begin{equation}
|R(\omega)|^{2}+|T(\omega)|^{2}=1.
\end{equation} 

Now let us discuss the WKB method developed in~\cite{Konoplya10a,Toshmatov16,Fernando17,Dey18}. If $ r_{0} $ is the value of $ r $ where the potential $ V(r) $ is the maximum, then depending on the relation between $ \omega $ and $V(r_{0})$, there are three cases to consider:

\begin{itemize}
	\item $ \omega^{2} << V(r_{0}) $. Here the transmission coefficient is close to zero and the reflection coefficient is almost equal to one.
	\item $ \omega^{2} >> V(r_{0}) $. Here the transmission coefficient is close to one and the reflection coefficient is almost equal to zero.
	\item $ \omega^{2} \sim V(r_{0}) $. We shall consider this case because the WKB approximation has high value of accuracy for  $ \omega^{2} \sim V(r_{0}) $.
\end{itemize}

In this approximation, the reflection coefficient is given by

\begin{equation}
R(\omega) = (1+e^{-2\pi i\alpha})^{-1/2},
\end{equation}

where $ \alpha $ is given by

\begin{equation}
\alpha = \frac{i(\omega^{2}-V_{0})}{\sqrt{-2V''_{0}}} - \Lambda_{2} - \Lambda_{3}.
\end{equation}

In the 3rd order WKB approximation $ \Lambda_{2} $ and $ \Lambda_{3} $ are given by

\begin{equation}
	\begin{aligned}
		& \Lambda_{2} = \frac{1}{\sqrt{-2V^{(2)}(r_{0})}} \Bigg[ \frac{1}{8} \Bigg( \frac{V^{(4)}(r_{0})}{V^{(2)}(r_{0})} \Bigg) (b^{2}+\frac{1}{4})  \\
		& \ \ \ \ \ \ \ - \frac{1}{288}\Bigg( \frac{V^{(3)}(r_{0})}{V^{(2)}(r_{0})} \Bigg)^{2}(7+60b^{2})  \Bigg], \\
		& \Lambda_{3} = \frac{n+\frac{1}{2}}{-2V^{(2)}(r_{0})} \Bigg[ \frac{5}{6912} \Bigg( \frac{V^{(3)}(r_{0})}{V^{(2)}(r_{0})} \Bigg)^{4} (188b^{2}+77)  \\
		& \ \ \ \ \ - \frac{1}{384}\Bigg( \frac{(V^{(3)})^{2}(r_{0})V^{(4)}(r_{0})}{(V^{(2)})^{3}(r_{0})} \Bigg)(51+100b^{2})  \\
		& +\frac{1}{2304} \Bigg( \frac{V^{(4)}(r_{0})}{V^{(2)}(r_{0})} \Bigg)^{2} (68b^{2}+67) -\frac{1}{288} \Bigg( \frac{V^{(6)}(r_{0})}{V^{(2)}(r_{0})} \Bigg) (4b^{2}+5) \\
		& \ \ \ \ \ + \frac{1}{288} \Bigg( \frac{V^{(3)}(r_{0})V^{(5)}(r_{0})}{(V^{(2)})^{2}(r_{0})} \Bigg) (28b^{2}+19)  \Bigg]
	\end{aligned}
\end{equation}

In these expressions $ b=n+\frac{1}{2} $, $ V^{n}(r_{0})=d^{n}V/dr_{*}^{n} $ at $ r = r_{0} $.

\begin{figure*}[t]
	\begin{center}
		\includegraphics[type=pdf,ext=.pdf,read=.pdf,width=5.5cm]{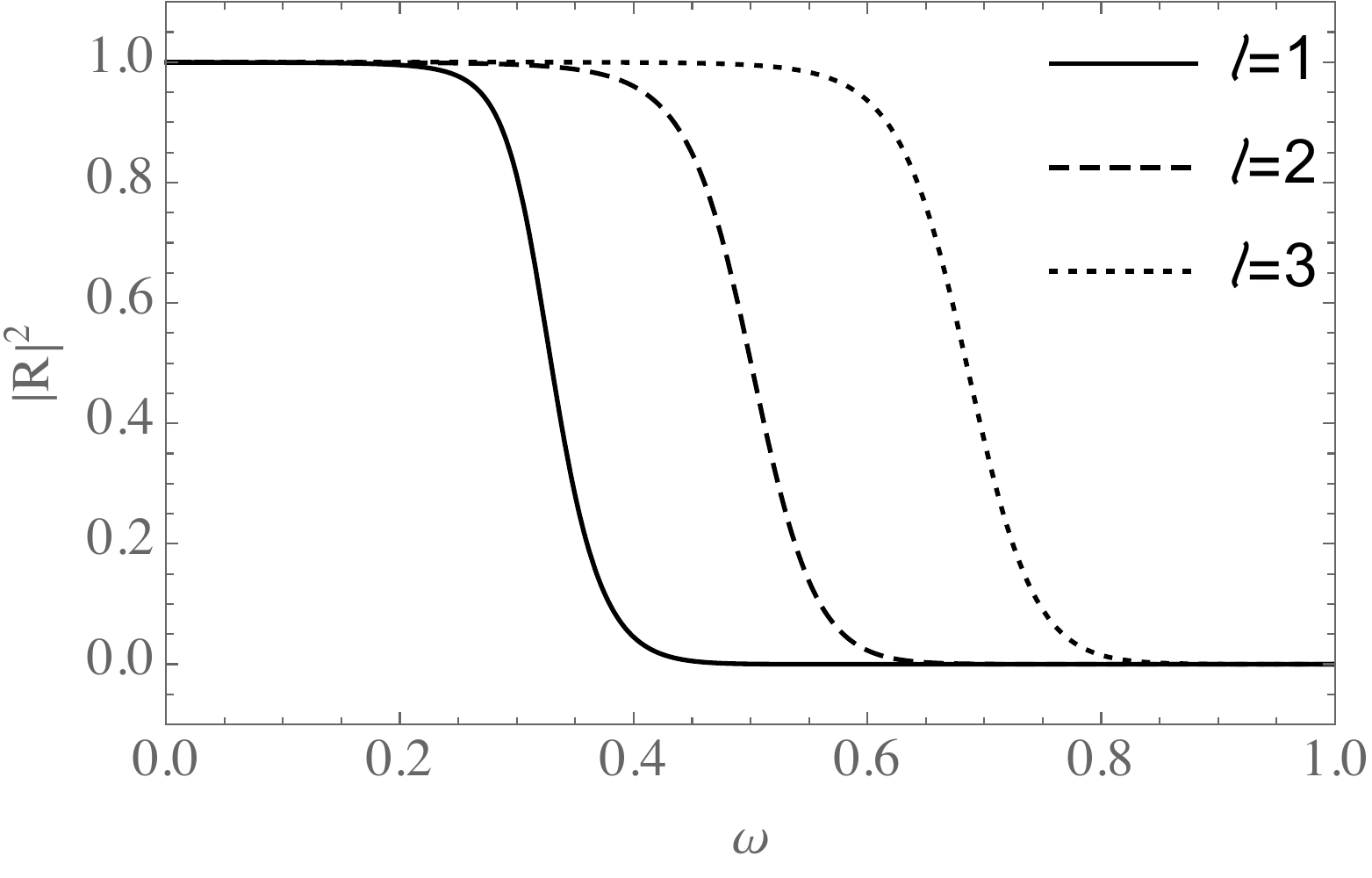}
		\includegraphics[type=pdf,ext=.pdf,read=.pdf,width=5.5cm]{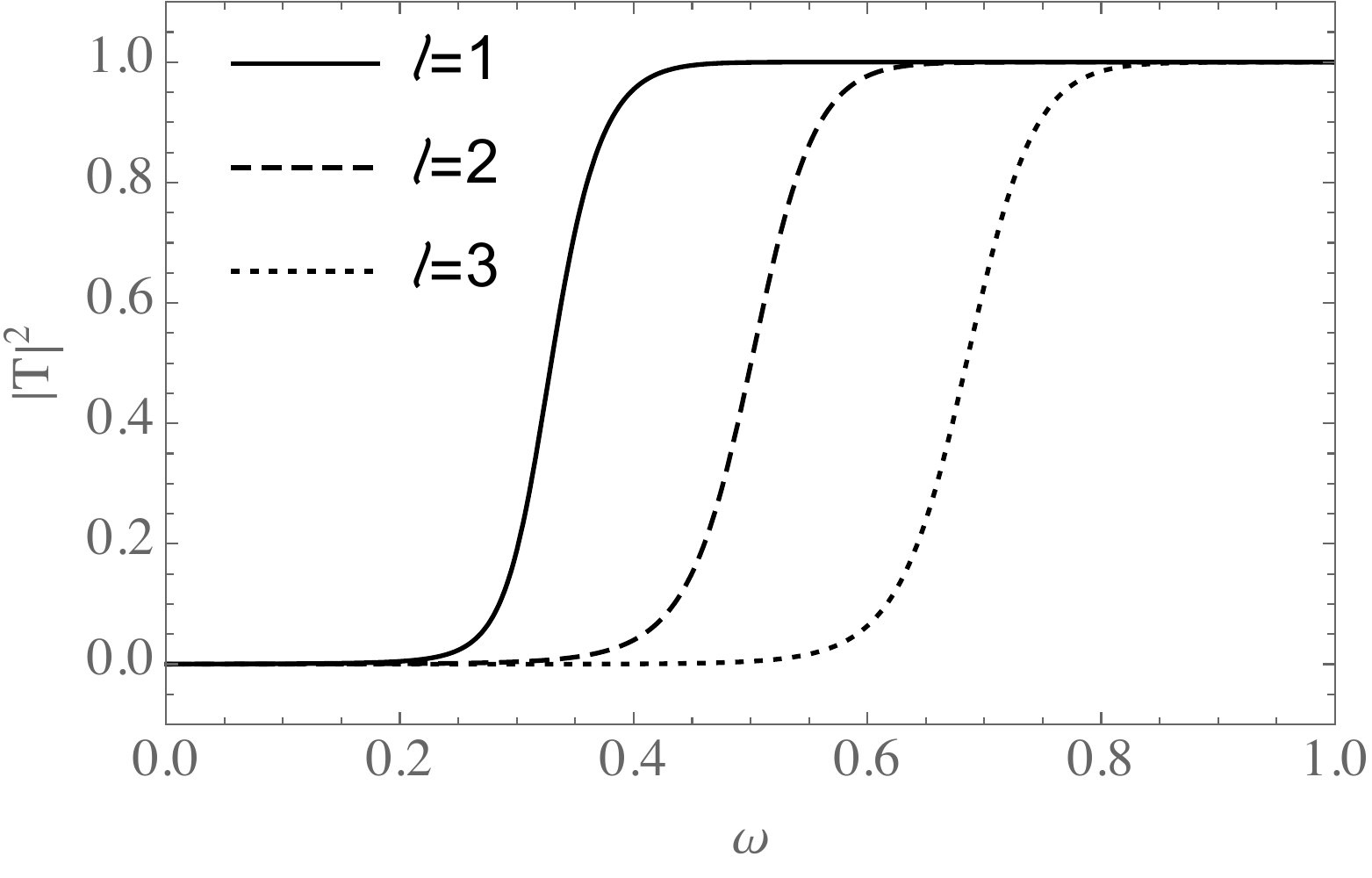}
		\includegraphics[type=pdf,ext=.pdf,read=.pdf,width=5.5cm]{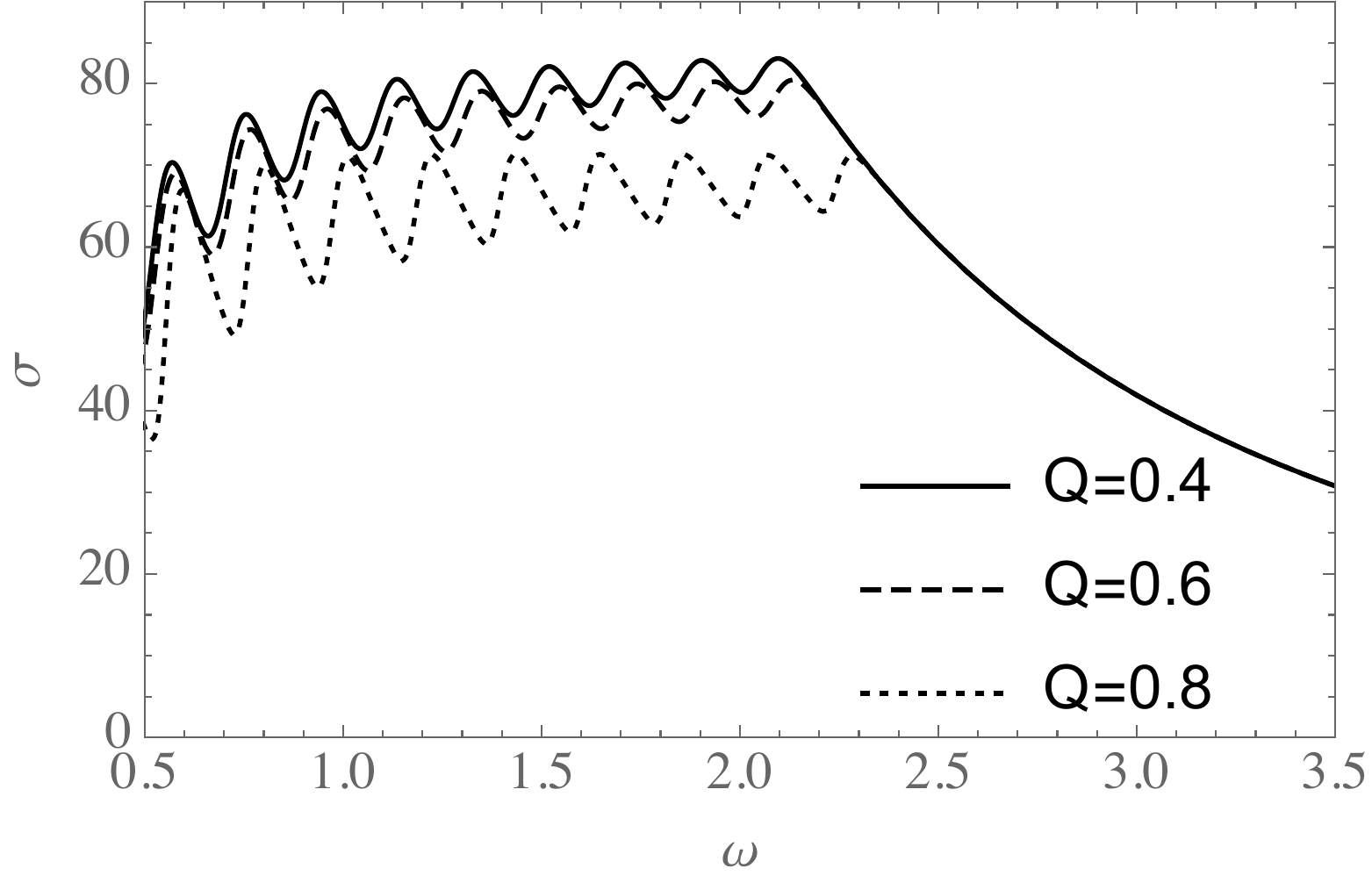}
	\end{center}
	\vspace{-0.5cm}
	\caption{ The left panel shows the dependence of $ |R(\omega)|^{2} $ on $ \omega $. The mid panel shows the dependence of $ |T(\omega)|^{2} $ or the greybody factor $ \gamma(\omega) $  on $ \omega $. Here $ M=1 $, $ Q=0.2 $, $ c=0.0001 $ and $ \omega_{q}=-2/3 $. The right panel shows the dependence of total absorption cross-section $ \sigma $ on $ \omega $ for $ Q = 0.4 $, $ Q = 0.6 $ and $ Q = 0.8 $.  \label{RT_sph} }
\end{figure*}

\begin{figure*}[t]
	\begin{center}
		\includegraphics[type=pdf,ext=.pdf,read=.pdf,width=5.5cm]{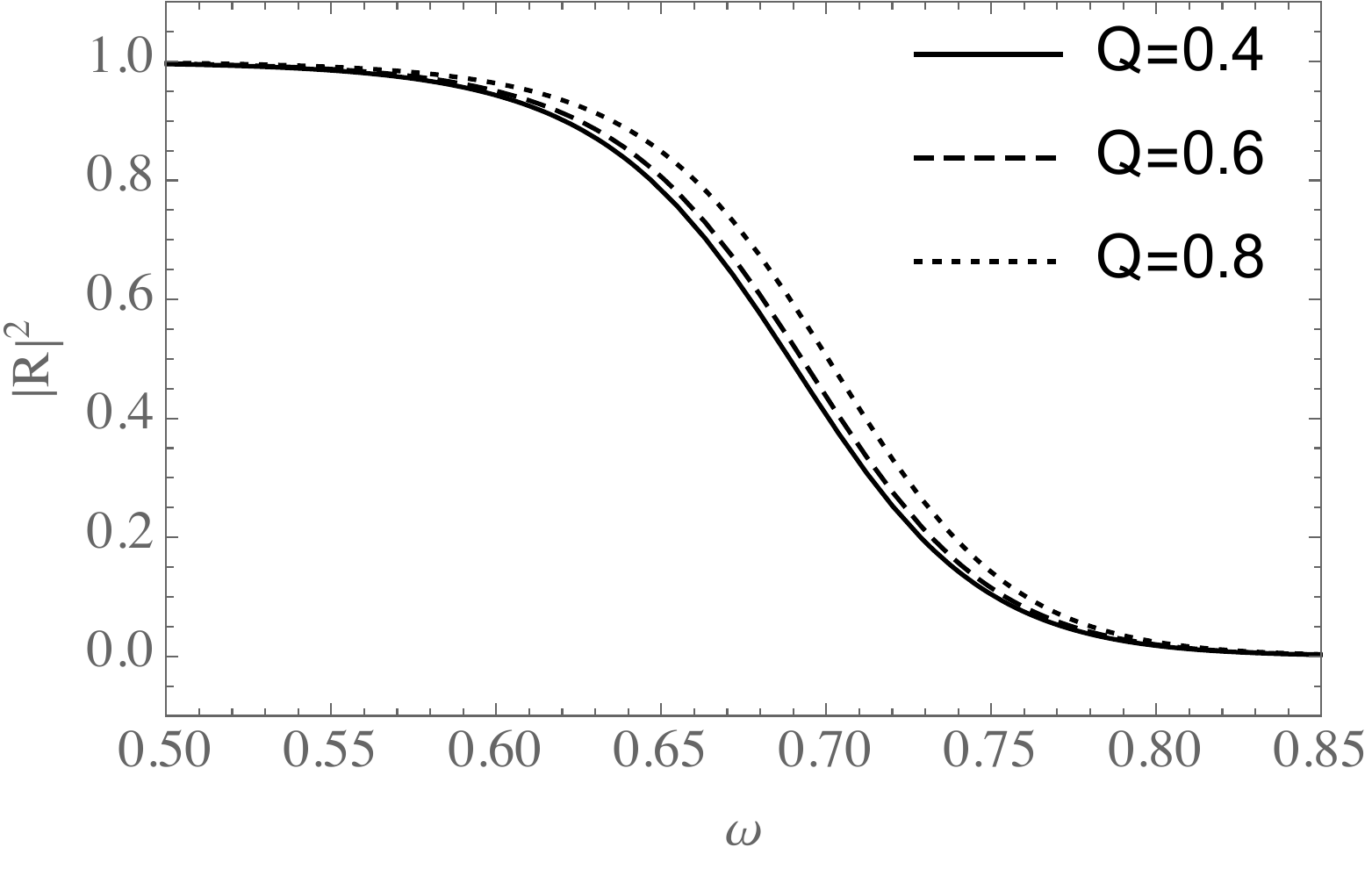}
		\includegraphics[type=pdf,ext=.pdf,read=.pdf,width=5.5cm]{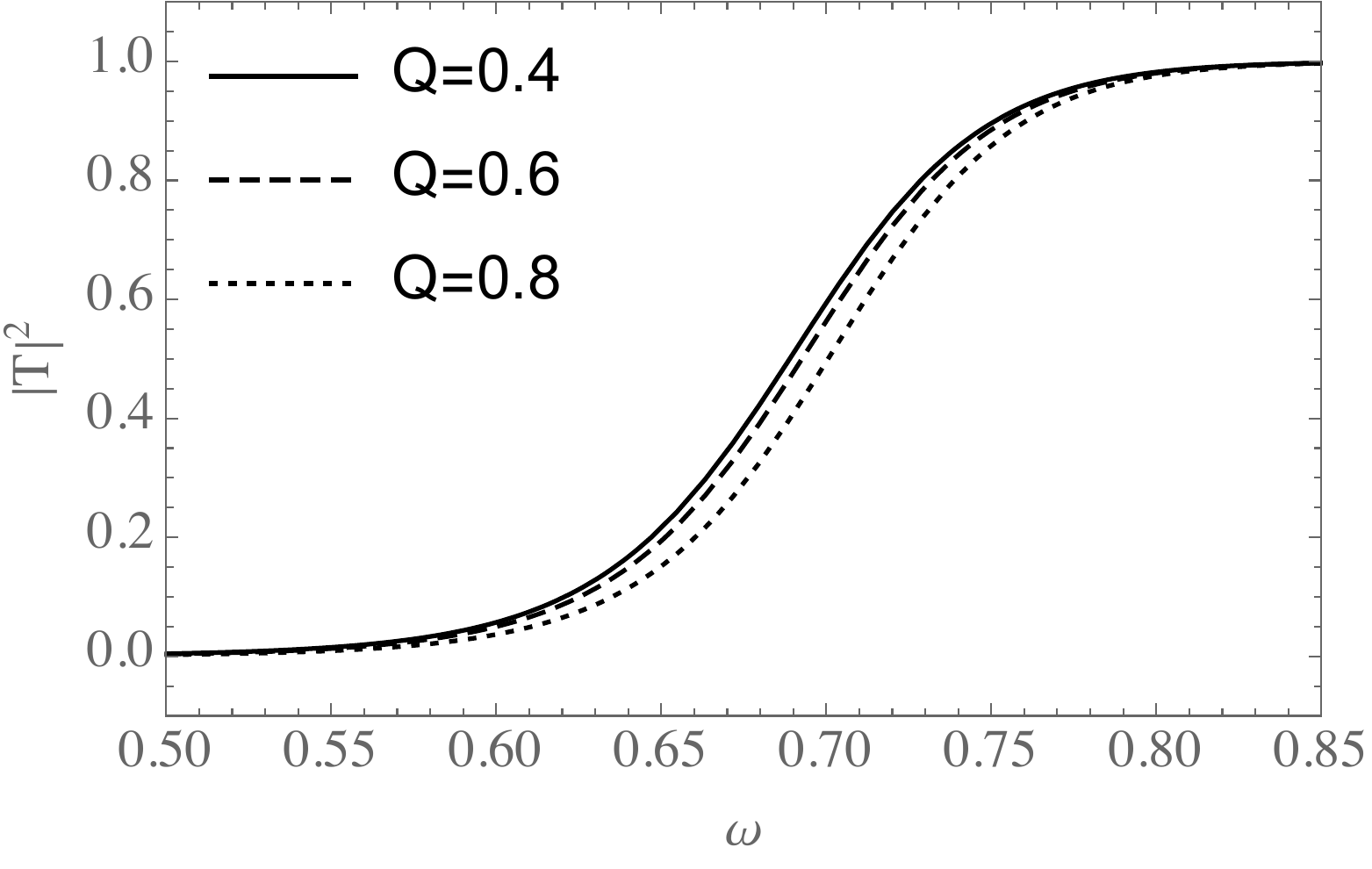}
		\includegraphics[type=pdf,ext=.pdf,read=.pdf,width=5.5cm]{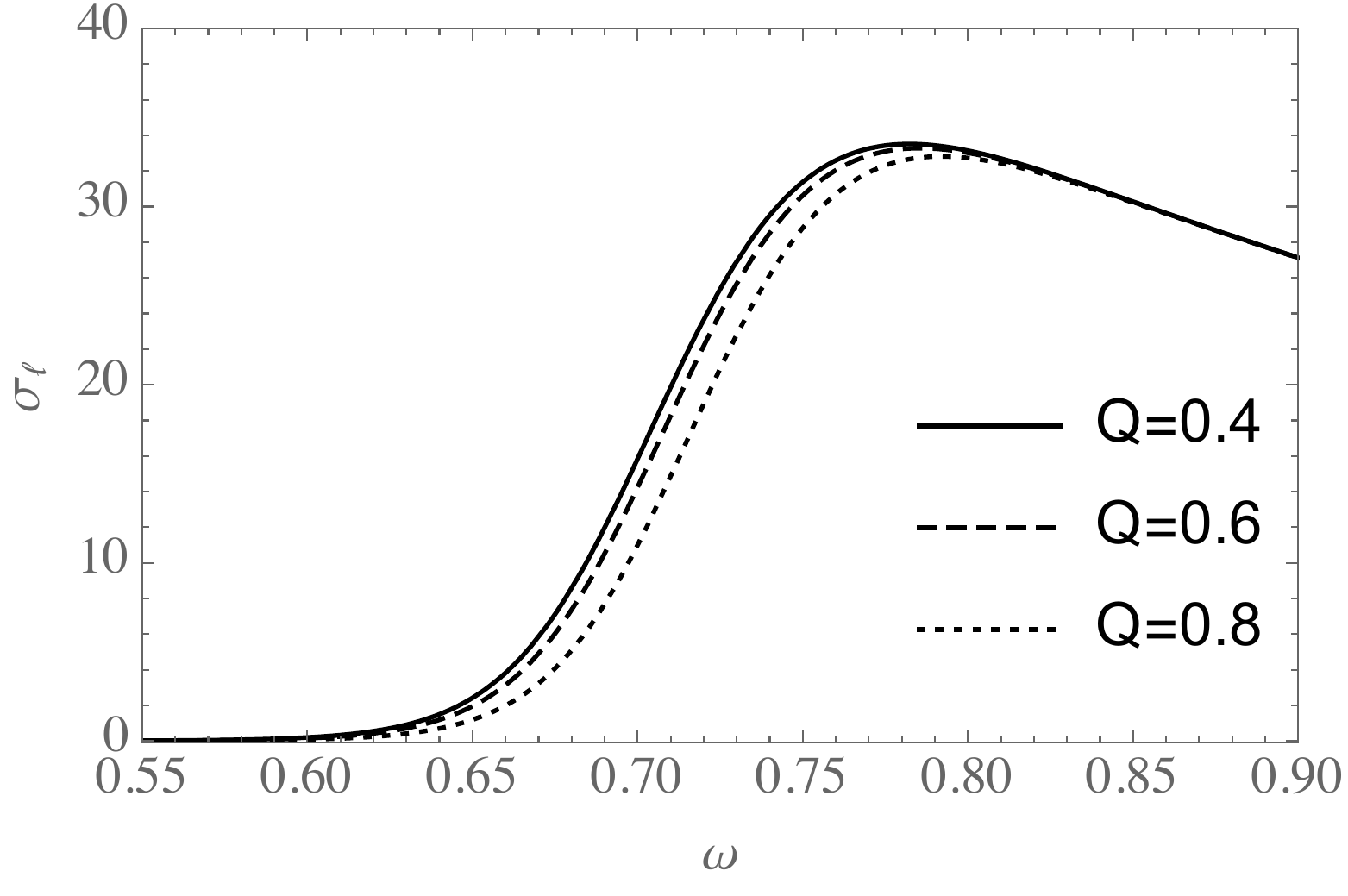}
	\end{center}
	\vspace{-0.5cm}
	\caption{ The left panel shows the dependence of $ |R(\omega)|^{2} $ on $ \omega $. The mid panel shows the dependence of $ |T(\omega)|^{2} $ or the greybody factor $ \gamma(\omega) $  on $ \omega $. The right panel shows the dependence of partial absorption cross-section $ \sigma_{l} $ on $ \omega $. Here $ M=1 $, $ l=3 $, $ c=0.0001 $ and $ \omega_{q}=-2/3 $.      \label{RT_charge} }
\end{figure*}

\begin{figure*}[t]
	\begin{center}
		\includegraphics[type=pdf,ext=.pdf,read=.pdf,width=5.5cm]{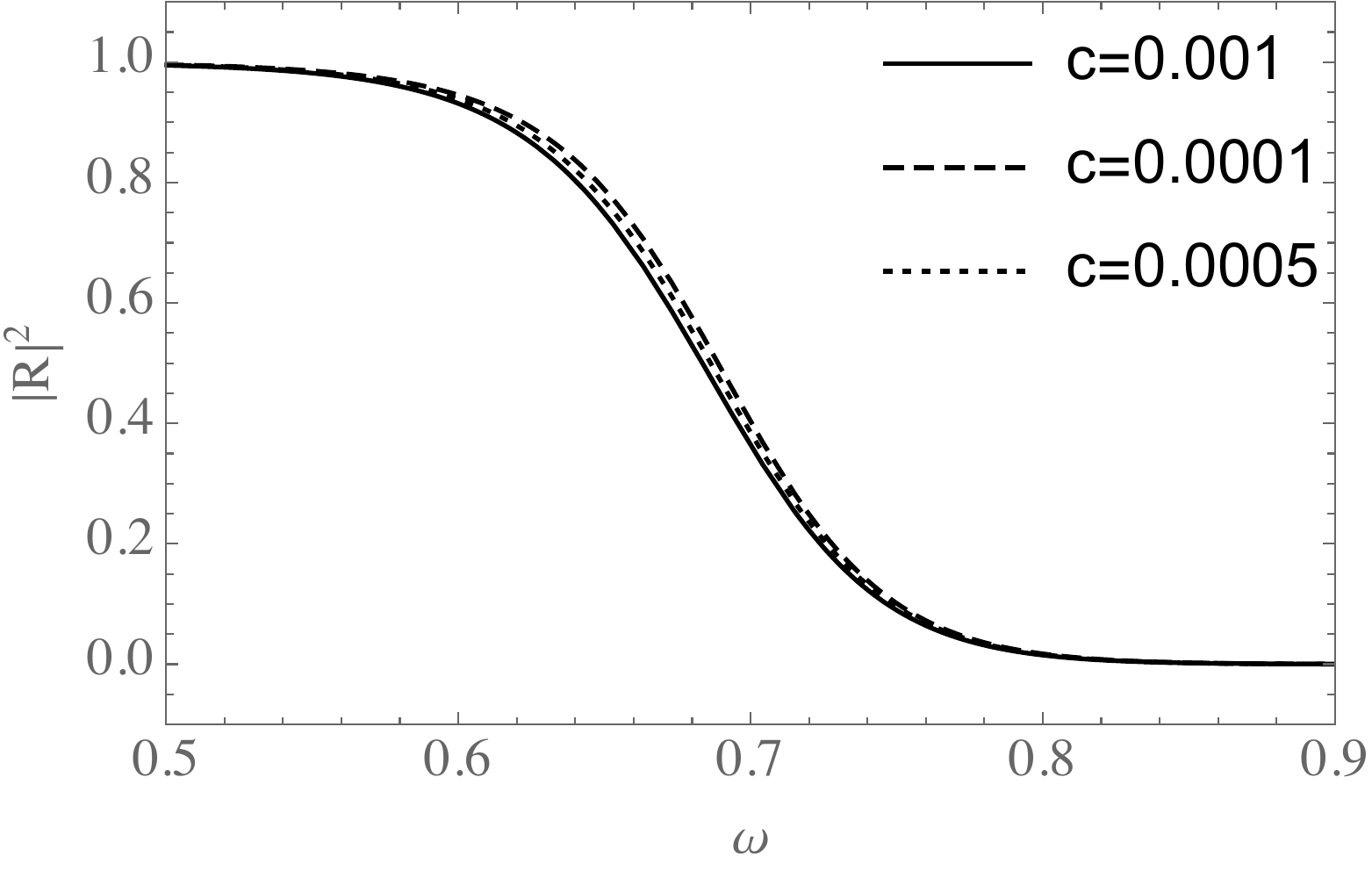}
		\includegraphics[type=pdf,ext=.pdf,read=.pdf,width=5.5cm]{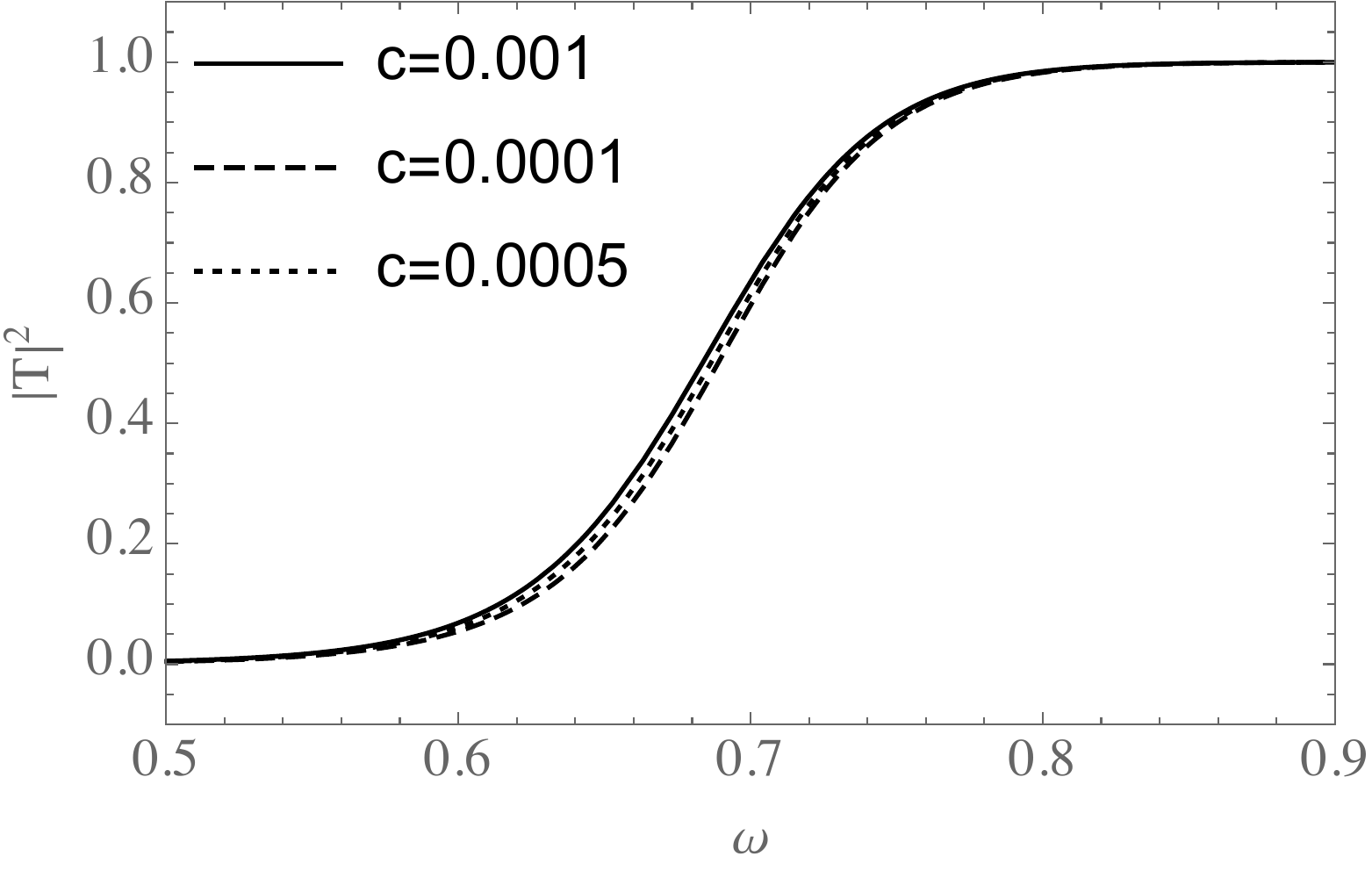}
		\includegraphics[type=pdf,ext=.pdf,read=.pdf,width=5.5cm]{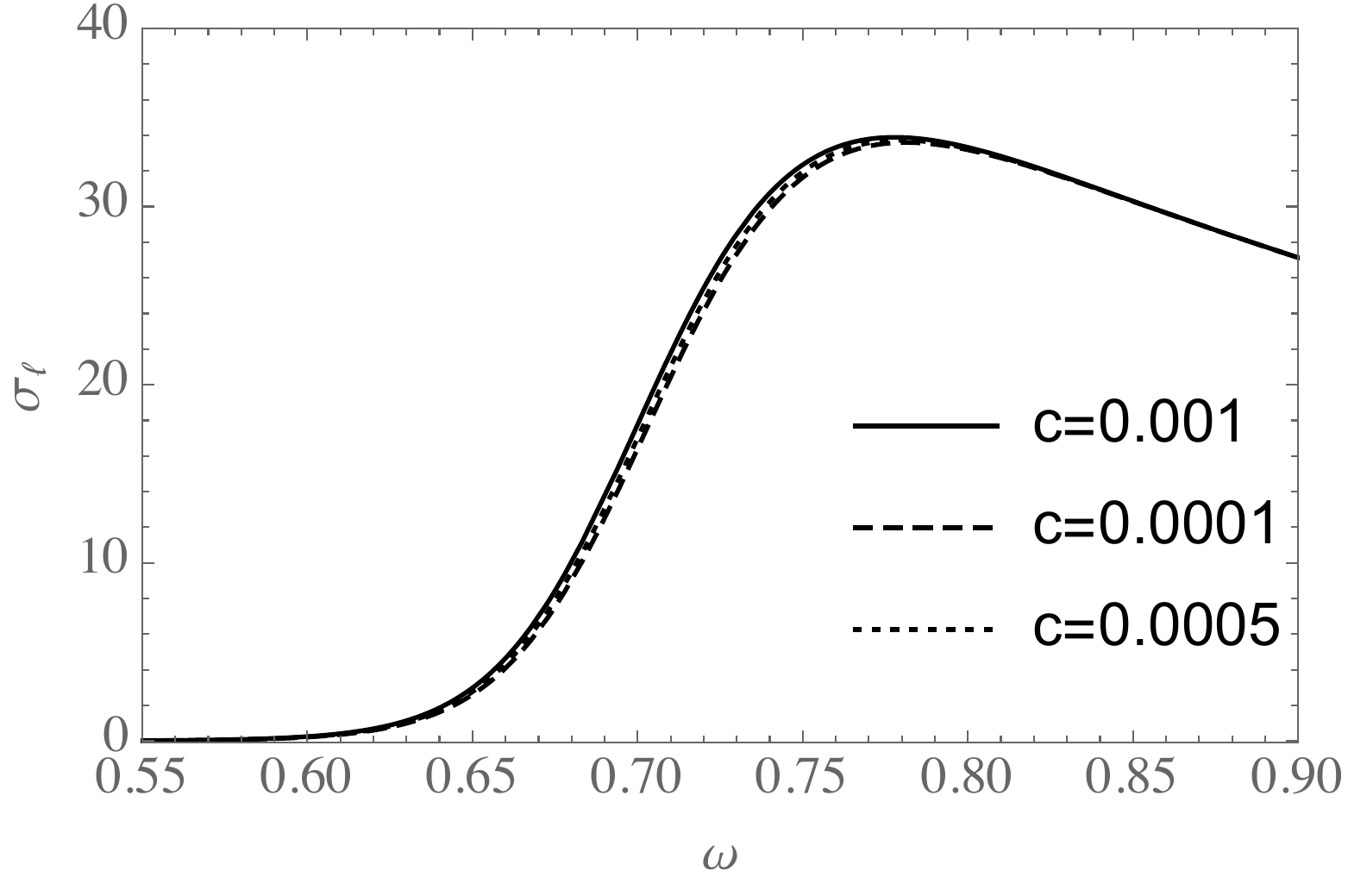}
	\end{center}
	\vspace{-0.5cm}
	\caption{  The left panel shows the dependence of $ |R(\omega)|^{2} $ on $ \omega $. The mid panel shows the dependence of $ |T(\omega)|^{2} $ or the greybody factor $ \gamma(\omega) $  on $ \omega $. The right panel shows the dependence of partial absorption cross-section $ \sigma_{l} $ on $ \omega $. Here $ M=1 $, $ Q=0.2 $, $ l=3 $ and $ \omega_{q}=-2/3 $.    \label{RT_c} }
\end{figure*}

\begin{figure}[t]
	\begin{center}
		\includegraphics[width=0.45\textwidth]{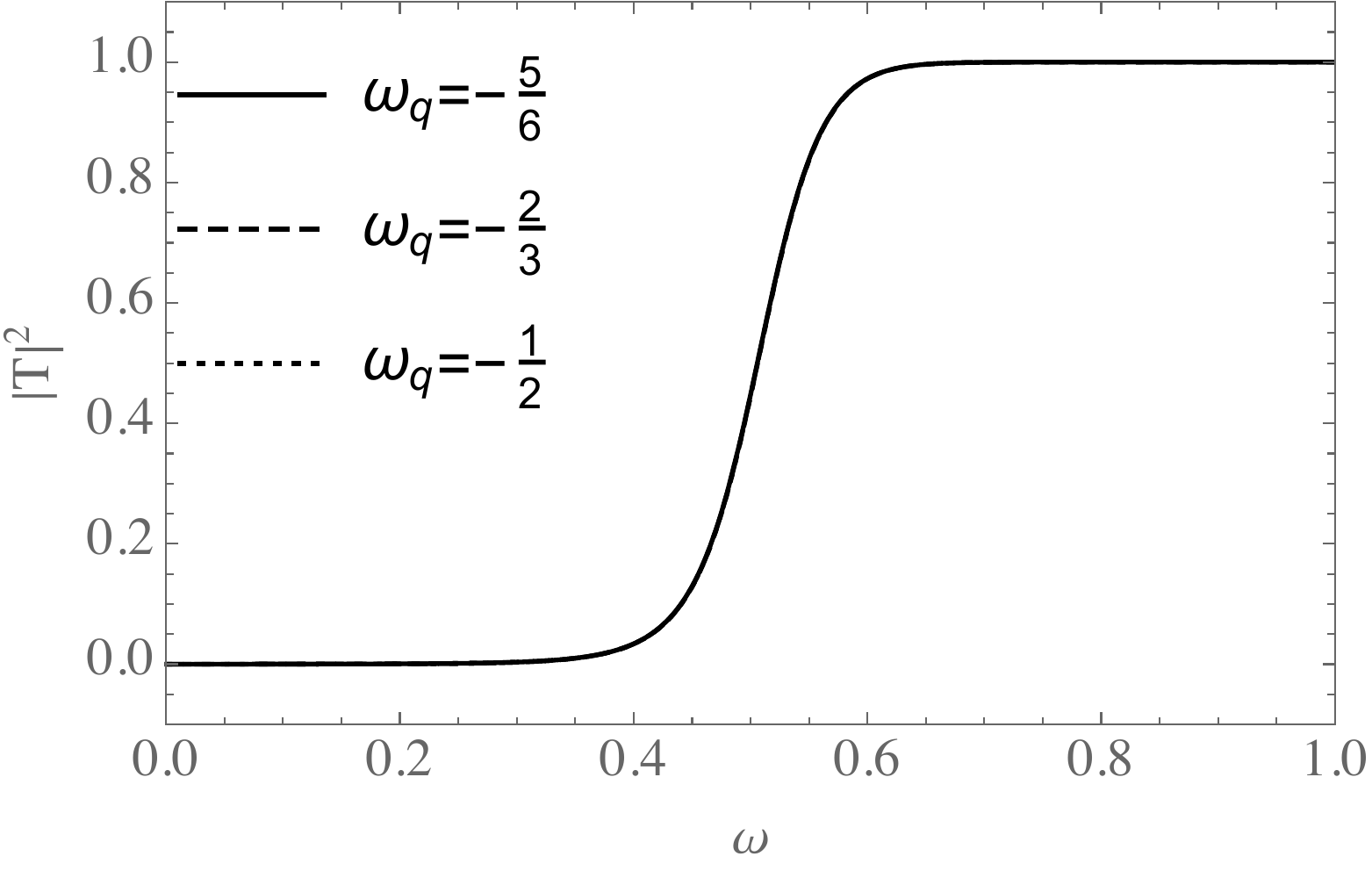}
	\end{center}
	\caption{Plot shows the dependence of greybody factor on $ \omega $ for different values of $ \omega_q $. The difference is indistinguishable here. \label{omega0}}
\end{figure}

\begin{figure*}[t]
	\begin{center}
		\includegraphics[type=pdf,ext=.pdf,read=.pdf,width=5.5cm]{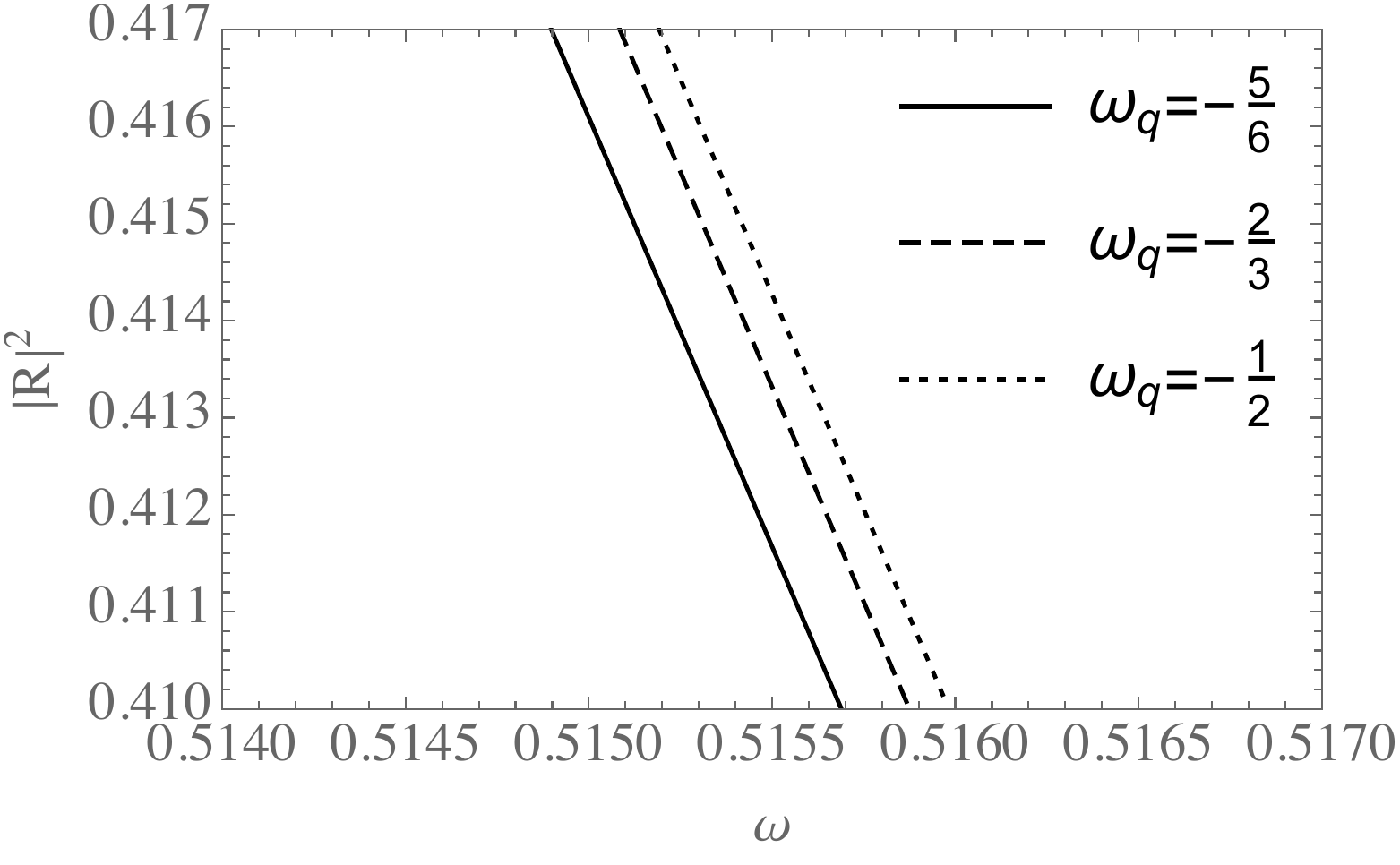}
		\includegraphics[type=pdf,ext=.pdf,read=.pdf,width=5.5cm]{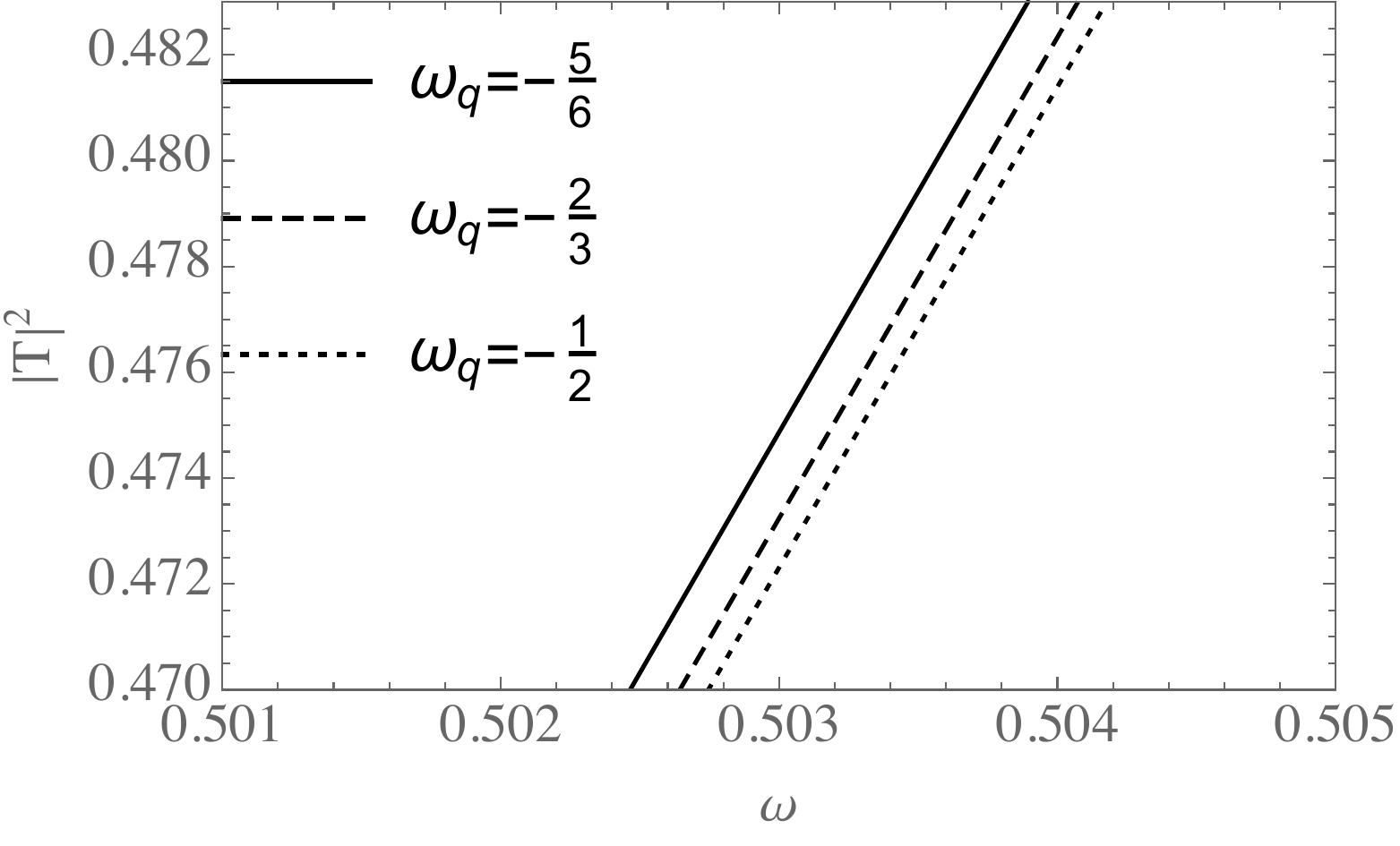}
		\includegraphics[type=pdf,ext=.pdf,read=.pdf,width=5.5cm]{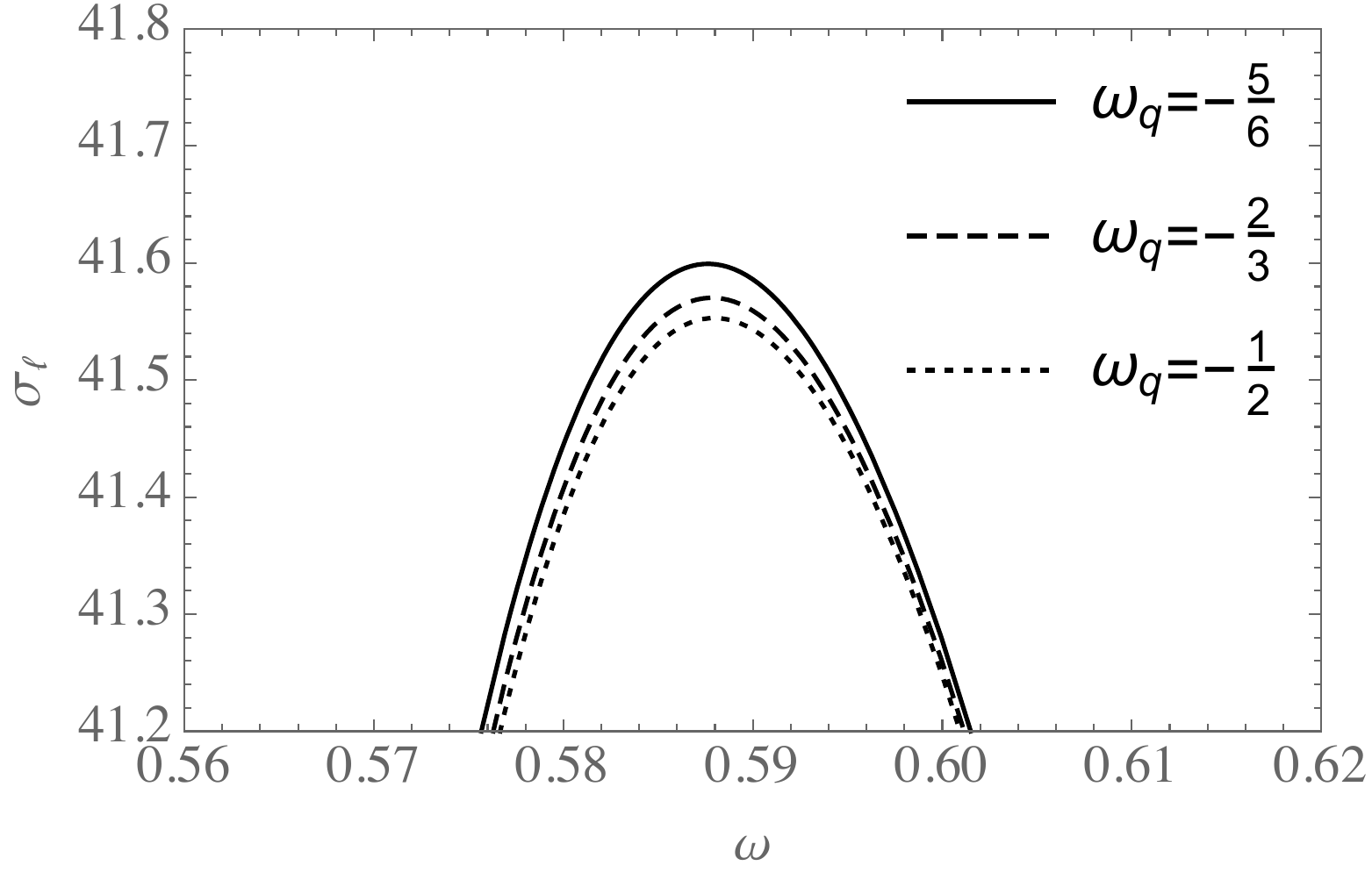}
	\end{center}
	\vspace{-0.5cm}
	\caption{  The left panel shows the dependence of $ |R(\omega)|^{2} $ on $ \omega $. The mid panel shows the dependence of $ |T(\omega)|^{2} $ or the greybody factor $ \gamma(\omega) $  on $ \omega $. The right panel shows the dependence of partial absorption cross-section $ \sigma_{l} $ on $ \omega $. Here $ M=1 $, $ Q=0.6 $, $ l=2 $ and $ c = 0.0001 $.    \label{omega1} }
\end{figure*}

In the left and middle panel of Fig.~\ref{RT_sph}, Fig.~\ref{RT_charge}, Fig.~\ref{RT_c} and Fig.~\ref{omega1}, we have plotted the dependence of reflection and absorption coefficient on the frequency of scalar perturbations, respectively. Fig.~\ref{RT_sph} shows the variation for different spherical harmonic indices $ l $, Fig.~\ref{RT_charge} shows the variation for different charges $ Q $, Fig.~\ref{RT_c} shows the variation for different quintessential parameter values $ c $ and Fig.~\ref{omega1} shows the variation for different values of quintessential parameter $ \omega_{q}$.

When the spherical harmonic index is varied, the transmission coefficient $ |T(\omega)|^{2} $ becomes smaller and hence the reflection coefficient $ |R(\omega)|^{2} $ becomes larger as seen in the mid and left panel of Fig.~\ref{RT_sph}. The Transmission coefficient $ |T(\omega)|^{2} $ decreases with the increase in magnetic charge and hence the reflection coefficient $ |R(\omega)|^{2} $ increases as can be seen in the mid and left panel of Fig.~\ref{RT_charge}. Similarly, $ |T(\omega)|^{2} $ increases with increase in quintessential parameter $ c $ and hence $ |R(\omega)|^{2} $ decreases as seen in the mid and left panel of Fig.~\ref{RT_c}. The change of transsmission and reflection coefficient is very small when we vary the quintessential paramter $ \omega_{q} $ as can be seen from Fig.~\ref{omega0}. In order to see the variation, we plot $ |R(\omega)|^{2} $, $ |T(\omega)|^{2} $ with higher resolution. We can see that transmission coefficient $ |T(\omega)|^{2} $ decreases for increase in quintessential parameter $ \omega_{q} $ and hence reflection coefficient $ |R(\omega)|^{2} $ increases.

\subsection{Absorption Cross-section}

In this subsection we shall discuss the partial absorption cross-section in the context of scalar perturbations for a non-linear magnetically charged black hole surrounded by quintessence. Partial absorption cross-section and total  absorption cross-section are defined as

\begin{equation}
\sigma_{l} = \frac{\pi(2l+1)}{\omega^{2}}|T(\omega)|^{2}.
\end{equation}
\begin{equation}
\sigma = \sum\limits_{l}\frac{\pi(2l+1)}{\omega^{2}}|T(\omega)|^{2}.
\end{equation}

We plot the variation of partial absorption cross-section $ \sigma_{l} $ with respect to the frequency of scalar perturbations in the right panels of Fig.~\ref{RT_charge}, Fig.~\ref{RT_c} and Fig.~\ref{omega1}. Fig.~\ref{RT_charge} shows the variation for different charges $ Q $ and Fig.~\ref{RT_c} shows the variation for different quintessential parameter values $ c $. Fig.~\ref{omega1} shows the variation of partial absorption with respect to the quintessential parameter $ \omega_{q} $. In these cases, the variation is extremely small. In the right panel of Fig.~\ref{RT_sph}, we have plotted the total absorption cross-section for different charges and for convenience we have summed over $ l = 1 $ to $ l = 10 $ modes to determine $ \sigma $. Total  absorption cross-section decreases for increasing value of charge parameter $ Q $. As the transmission coefficient attains the value of one at some critical value of $ \omega $, the total absorption crosssection falls off as $ 1/\omega^2 $ regardless of the black hole parameters. Therefore we observe the fall-off region in this particular plot.

\section{Summary and Conclusion\label{conclusion}}

In this paper, we have focused on scalar perturbations of a nonlinear magnetic charged black hole surrounded by quintessence. First, we numerically calculated the time evolution of scalar perturbations around the considered black hole spacetime and we found that perturbations with higher spherical harmonic index $ l $ live longer. We also saw the behaviour of perturbations with changing cosmological constant parameter $ c $ and we note that perturbations with the higher value of $ c $ becomes unstable. The behaviour of perturbation for different values of charge parameter is also studied and reported in the manuscript. 

We have used the 6th order WKB method to calculate the quasinormal frequencies of scalar perturbations for a nonlinear magnetic charged black hole surrounded by quintessence. We studied the dependence of quasinormal frequencies on charge $ Q $, spherical harmonic index $ l $ and cosmological constant parameter $ c $. We see that the real part of quasinormal frequency increases with increase in charge $ Q $ for both fundamental ($ n = 0 $) and first overtone mode ($ n = 1 $) but the magnitude of the imaginary part of the frequency decreases. The magnitude of the imaginary part of the frequency decreases with increase in charge for different values of $ c $. The real part of the quasinormal frequencies increase monotonically with respect to $ l $ but the variation with respect to $ c $ is extremely small and the lines overlap on the plot. The magnitude of the imaginary frequencies decreases with an increase in $ l $ and $ c $.  

We have further studied the greybody factors $ \gamma(\omega) $ and the partial absorption cross-section $ \sigma_{l} $ of scalar perturbations for a nonlinear magnetic charged black hole surrounded by quintessence. We studied the dependence of these quantities for different parameters of the black hole spacetime.  The transmission coefficient $ |T(\omega)|^{2} $, or the greybody factor $ \gamma(\omega) $, becomes smaller and hence reflection coefficient $ |R(\omega)|^{2} $ becomes larger with increase in $ l $. The transmission coefficient $ |T(\omega)|^{2} $ decreases with the increase in charge and hence the reflection coefficient $ |R(\omega)|^{2} $ increases. Similarly, $ |T(\omega)|^{2} $ increases with increase in quintessential parameter $ c $ and hence $ |R(\omega)|^{2} $ decreases but the change is too small. We also investigated the effect of changing quintessential parameter $ \omega_{q} $ on transmission and reflection coefficient and partial absorbtion cross-section and we found that the effect of $ \omega_{q} $ on these quantities are very small. 

For future directions, it would be interesting to study the electromagnetic and gravitational perturbations for the nonlinear magnetic charged black hole surrounded by quintessence. We hope to understand the behavior of these perturbations with respect to the cosmological constant parameter $ c $.

\begin{acknowledgments}
	
	The authors thank Dimitry Ayzenberg and Askar Abdikamalov for helpful discussion.
	 This work was supported by the National Natural Science Foundation of China (Grant No.~U1531117) and Fudan University (Grant No.~IDH1512060). 
	 H.C. also acknowledges support from the China Scholarship Council (CSC), grant No.~2017GXZ019020.  
	 The research is supported in part by Grant No. VA-FA-F-2-008 and No.YFA-Ftech-2018-8 of the Uzbekistan Ministry for Innovation Development, by the Abdus Salam International Centre for Theoretical Physics through Grant No. OEA-NT-01 and by Erasmus+ exchange grant between Silesian University in Opava and National University of Uzbekistan. A.A. thanks the Nazarbayev University for the hospitality. 
	 
\end{acknowledgments}

\bibliography{gravreferences}

\end{document}